\documentclass[a4paper,fleqn]{cas-dc}

\usepackage[authoryear]{natbib}

\usepackage{algorithm}
\usepackage{algpseudocode}
\usepackage{tikz}
\usepackage{pgfplots}
\usepackage{caption}
\usepackage{subcaption}
\usetikzlibrary{arrows.meta, positioning, shapes.geometric, fit, calc, backgrounds}
\pgfplotsset{compat=1.18}

\newcommand{\Pfeas}{P_{\mathrm{feas}}}
\newcommand{\Popt}{P_{\mathrm{opt}}}
\newcommand{\Efeas}{\mathbb{E}[C\,|\,\mathrm{feas}]}
\newcommand{\Dtwo}{D_{2q}}

\def\tsc#1{\csdef{#1}{\textsc{\lowercase{#1}}\xspace}}
\tsc{WGM}
\tsc{QE}
\tsc{EP}
\tsc{PMS}
\tsc{BEC}
\tsc{DE}

\begin{document}
\let\WriteBookmarks\relax
\def\floatpagepagefraction{1}
\def\textpagefraction{.001}

\shorttitle{Hardware-Efficient Quantum Optimization for Transportation Networks via Compressed Adiabatic Evolution}

\shortauthors{T. Azfar, R. Ke, S. He, C. Wang, J. Holguín-Veras}

\title [mode = title]{Hardware-Efficient Quantum Optimization for Transportation Networks via Compressed Adiabatic Evolution}     



%
\author[]{Talha Azfar}[type=editor,
                        auid=000,bioid=1,
                        orcid=0000-0002-1293-5036]




\credit{Conceptualization of this study, Methodology, Software}

\affiliation[]{organization={Rensselaer Polytechnic Institute},
    addressline={110 8th Street}, 
    city={Troy, NY},
    citysep={}, 
    postcode={12180}, 
    country={USA}}

\author[]{Ruimin Ke}[type=editor,
                        auid=000,bioid=1,
                        orcid=0000-0001-9139-6765]

\cormark[1]
\ead{ker@rpi.edu}
\credit{Data curation, Writing - Original draft preparation}
                        
\author[]{Sean He}[type=editor,
                        auid=000,bioid=1,
                        orcid=0000-0002-1633-4748]

\author[]{Cara Wang}[type=editor,
                        auid=000,bioid=1,
                        orcid=0000-0001-5819-0212]

\author[]{Jos\'e Holguín-Veras}[type=editor,
                        auid=000,bioid=1,
                        orcid=0000-0001-8118-9383]




\cortext[cor1]{Corresponding author}



\begin{abstract}
Transportation systems such as urban logistics, vehicle routing, and infrastructure planning require solving large-scale combinatorial optimization problems under complex constraints. Problems such as the vehicle routing problem (VRP), traveling salesman problem (TSP), and facility location problem (FLP) involve large discrete search spaces and the need to generate multiple feasible solutions in real time.
In this work, we develop a hardware-grounded hybrid quantum optimization framework that uses Approximate Quantum Compilation (AQC) to compress early segments of digitized adiabatic evolution into shallow circuits. The compressed prefix is combined with variational layers, enabling a systematic study of how initialization, circuit depth, and expressivity interact on near-term quantum hardware. All experiments are performed on an IBM gate-based quantum computer, and circuits are evaluated as stochastic generators of candidate transportation plans.
Results show that moderate prefix compression reduces two-qubit gate depth while maintaining or improving feasible solution discovery, particularly for routing problems. These benefits depend on compatibility between the compressed prefix and the variational ansatz: while standard QAOA effectively leverages AQC initialization, linear-chain QAOA shows limited improvement.
Overall, this work demonstrates that hybrid AQC-QAOA methods provide a practical pathway for hardware-efficient quantum optimization, positioning quantum algorithms as candidate generators within transportation decision-making workflows.



\end{abstract}



\begin{keywords}
Quantum Computing \sep Facility Location \sep Vehicle Routing \sep Traveling Salesman Problem \sep Optimization \sep QAOA
\end{keywords}

\maketitle

\section{Introduction}

Modern transportation and logistics systems depend critically on the ability to solve large-scale combinatorial optimization problems that arise across multiple levels of planning and operations. Many of the key decisions required to manage these systems involve selecting discrete actions over complex network structures, where the number of feasible configurations grows combinatorially with system size. Examples of such decisions include the strategic placement of infrastructure such as warehouses or distribution centers, the tactical design of vehicle routes serving geographically distributed demand, and the real-time dispatch of vehicles and resources in response to dynamic conditions. These decisions must typically be made under resource constraints, network connectivity limitations, and operational requirements that significantly increase the complexity of the resulting optimization models.

A number of well-established combinatorial optimization formulations capture these transportation decision problems. The Vehicle Routing Problem (VRP) models the task of determining efficient routes for a fleet of vehicles that must serve a set of customers while minimizing travel cost or time \citep{clarke1964scheduling}. The Traveling Salesman Problem (TSP) represents a simplified single-vehicle routing problem in which a tour must visit each location exactly once before returning to the starting point \citep{junger1995traveling}. The Facility Location Problem (FLP) addresses strategic infrastructure planning by determining which candidate facilities should be opened and how customers should be assigned to them in order to minimize combined facility and service costs \citep{aardal1996two}. Although these models provide useful abstractions of real-world transportation planning problems, they remain computationally challenging due to their combinatorial structure, and their difficulty increases rapidly as the number of locations, vehicles, and operational constraints grows. In practice, transportation analytics therefore blends exact optimization (e.g., mixed-integer programming) with heuristics and metaheuristics that trade optimality guarantees for computational tractability \citep{laporte2007you}.

The increasing scale and complexity of modern transportation systems further amplify these computational challenges. Urban freight systems involve interactions among logistics providers, infrastructure constraints, regulatory policies, and stochastic demand patterns. Evaluating alternative system configurations frequently requires exploring large combinatorial search spaces while simultaneously accounting for operational feasibility constraints. These characteristics motivate the exploration of new computational paradigms capable of complementing classical optimization methods.

Quantum computing has recently emerged as a potential framework for solving certain classes of combinatorial optimization problems. Many discrete optimization models---including routing, scheduling, and assignment problems---can be expressed as Quadratic Unconstrained Binary Optimization (QUBO) formulations or equivalently as Ising Hamiltonians \citep{lucas2014ising}, enabling them to be represented within quantum optimization algorithms. Two prominent approaches are adiabatic quantum computing \citep{farhi2000adiabatic} and the Quantum Approximate Optimization Algorithm (QAOA) \citep{farhi2014qaoa}. Both frameworks aim to prepare quantum states whose measurement yields high-quality solutions to the underlying optimization problem.

Despite these promising theoretical developments, practical deployment of quantum optimization for transportation systems faces significant challenges. Current quantum devices operate in the Noisy Intermediate-Scale Quantum (NISQ) regime, where circuit depth, qubit connectivity, and noise levels impose strict limits on executable algorithms \citep{preskill2018quantum}. In particular, digitized implementations of adiabatic evolution often require deep circuits containing many sequential two-qubit gates, while variational algorithms such as QAOA introduce high-dimensional parameter optimization problems that can be difficult to train. Consequently, designing hardware-efficient quantum algorithms remains a central research challenge. 

A key observation motivating this work is that the computational difficulty of adiabatic evolution is not uniform throughout the evolution process.  The spectral structure of the system Hamiltonian typically exhibits distinct regimes characterized by different energy-gap behaviors. Early and late portions of the evolution often possess relatively large spectral gaps (difference between the lowest and second lowest eigenvalues), making the dynamics more robust to perturbations, while intermediate regions may contain narrow spectral gaps associated with avoided crossings that govern the difficulty of reaching the ground state. In fact, the running time of the adiabatic computation is directly determined by the minimal spectral gap \citep{aharonov2008adiabatic}. Although directly measuring spectral gaps on current quantum hardware is generally infeasible, the presence of such gap-critical regions often manifests indirectly through increased state mixing and broader cost distributions in sampled measurement outcomes. Unlike quantum annealers, which perform analog optimization via continuous-time evolution, our approach targets gate-based quantum computers, where annealing dynamics are discretized into circuits that must operate under depth and connectivity constraints.

In this work, we therefore adopt a hardware-accessible diagnostic based on the statistical properties of measured solution distributions. In particular, moments of the sampled objective function, such as energy variance of the cost distribution and susceptibility of the Hamiltonian, serve as empirical indicators of state mixing during the evolution. These quantities can be computed directly from measured bitstring samples without requiring full state tomography or explicit Hamiltonian diagonalization. Peaks in these quantities can indicate regions where the evolving quantum state spreads across multiple nearby energy levels, providing a practical proxy for gap-sensitive regions of the optimization landscape.

\begin{figure*}[t]
\centering
\usetikzlibrary{fit,backgrounds}

\begin{tikzpicture}[
step/.style={draw, minimum width=0.85cm, minimum height=0.7cm, font=\scriptsize},
label/.style={font=\small},
box/.style={draw, rounded corners=3pt, minimum width=3.8cm, minimum height=0.9cm, align=center, fill=blue!12},
var/.style={draw, minimum width=0.9cm, minimum height=0.7cm, fill=orange!15, font=\scriptsize},
prefixgroup/.style={draw, rounded corners=4pt, fill=blue!12, inner xsep=10pt, inner ysep=8pt},
arrow/.style={-Latex, thick}
]

\node[label] at (5,2.1) {\textbf{Digitized quantum annealing baseline}};

\foreach \i in {0,...,9}
{
\node (u\i) at (\i*1.0,1.2) {$U_{\i}$};
}

\begin{pgfonlayer}{background}
\node[prefixgroup, fit=(u0)(u1)(u2)(u3)] {};
\end{pgfonlayer}

\node[label] at (1.8,0.5) {early evolution};
\node[label] at (5.0,0.5) {gap-critical region};
\node[label] at (8.2,0.5) {late evolution};

\draw[arrow] (5,0.2) -- (5,-0.5);

\node[draw, rounded corners=4pt, fill=yellow!20, minimum width=4.2cm, minimum height=0.9cm, align=center]
at (5,-1)
{Compress large-gap region\\variationally treat gap-critical region};

\node[label] at (5,-1.9) {\textbf{Proposed hybrid hardware-efficient framework}};

\node[box] (prefix) at (1.8,-3) {Compressed prefix\\(AQC)};

\foreach \j/\x in {1/4.6,2/5.7,3/6.8,4/7.9,5/9.0}
{
\node[var] at (\x,-3) {$Q_{\j}$};
}

\node[label] at (6.8,-3.9) {QAOA or LC-QAOA layers};

\end{tikzpicture}

\caption{
Hybrid hardware-efficient quantum optimization architecture.
A digitized quantum annealing circuit is decomposed into segments with different computational roles.
The early portion of the evolution is compressed using approximate quantum compilation (AQC),
while the remaining gap-critical portion is replaced by structured variational layers
such as QAOA or LC-QAOA. This design reduces circuit depth while preserving optimization capability.
}
\label{fig:hybrid_architecture}

\end{figure*}
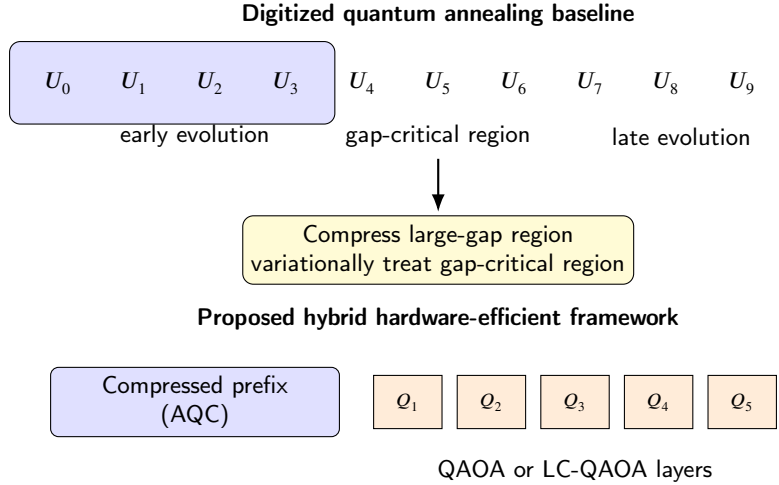

Building on this observation, this paper proposes hardware-efficient quantum architectures for transportation optimization by comparing circuit families that are implementable today. The proposed approach decomposes digitized adiabatic evolution into two components. First, the early portion of the evolution is compressed using Approximate Quantum Compilation (AQC) \citep{robertson2025approximate,qiskit-addon-aqc-tensor}, which replaces multiple Trotterized time-evolution steps with a shallower circuit approximation. Second, the remaining portion of the evolution is replaced with structured variational circuits based on either standard QAOA or a Linear-Chain QAOA (LC-QAOA) architecture designed to reduce two-qubit gate depth under hardware connectivity constraints. The concept is shown in Figure \ref{fig:hybrid_architecture}.

The goal of this study is to analyze the computational tradeoffs associated with different quantum algorithm architectures rather than computational advantage. Specifically, the paper investigates how circuit compression, variational flexibility, and hardware connectivity interact to influence three key properties: circuit depth, solution feasibility probability, and optimization landscape behavior. We systematically study the effect of AQC-based prefix compression across FLP, VRP, and TSP instances, evaluating performance not only in terms of objective value, but also through transportation-relevant metrics such as feasible solution rates and diversity of candidate solutions. This perspective reflects the practical role of optimization in transportation systems, where generating multiple feasible alternatives is often as important as identifying a single optimal solution.

Importantly, all experiments in this work are executed on real quantum hardware, allowing us to evaluate algorithm performance under practical noise and connectivity constraints. We deliberately focus on small, tractable problems to ensure that two-qubit depth on quantum hardware does not pose an immediate bottleneck to our analysis. Our results reveal several key insights. First, AQC prefix compression can significantly reduce circuit depth while preserving feasible-solution discovery, but its effectiveness depends strongly on the structure of the underlying problem formulation. Assignment-based models such as TSP exhibit more predictable depth reductions, while routing and siting formulations show more irregular compilation behavior. Second, moderate compression can improve or maintain feasible-solution generation in hybrid AQC+QAOA circuits, particularly for VRP instances, supporting the use of quantum algorithms as candidate generators under hardware constraints.

The proposed framework therefore serves as a computational roadmap for designing hardware-aware quantum optimization algorithms for transportation applications. By systematically analyzing the interaction between algorithm design and hardware constraints, this work aims to clarify how emerging quantum optimization methods may eventually integrate with classical transportation analytics in future hybrid computing environments. Overall, this paper positions hybrid quantum optimization not as a standalone solver, but as a hardware-aware candidate-generation framework for transportation systems. By explicitly analyzing the tradeoffs between circuit depth, feasibility, and solution quality, we provide guidance on how to co-design problem formulations, initialization strategies, and variational circuits for practical quantum optimization in transportation applications.

The main contributions of this work are summarized as follows:
\begin{itemize}
\item \textbf{Hybrid hardware-efficient quantum optimization framework:}
We develop and evaluate a hardware-aware hybrid quantum optimization framework that combines AQC-based prefix compression with variational circuits for transportation problems, including FLP, VRP, and TSP.

\item \textbf{Spectral-structure-aware design:}
We highlight how the spectral-gap structure of adiabatic evolution can inform algorithm design, enabling selective compression of evolution segments that are less sensitive to perturbations. Distribution based metrics derived from sampled objective values are introduced as practical proxies for identifying gap-sensitive regions of the adiabatic path.

\item \textbf{Structured variational circuit comparison:}
We provide a systematic analysis of the tradeoffs between circuit depth, feasible-solution discovery, and solution quality across the $(m,p)$ parameter grid, highlighting the role of problem formulation and encoding structure in determining compilation efficiency.

\item \textbf{Implications for transportation optimization research:}
We show that the benefits of AQC-based initialization are not universal across variational architectures, establishing the importance of compatibility between initialization and variational ans\"atz design. The results provide insight into how hybrid quantum-classical computing architectures may be developed for large-scale transportation optimization problems in the future.
\end{itemize}

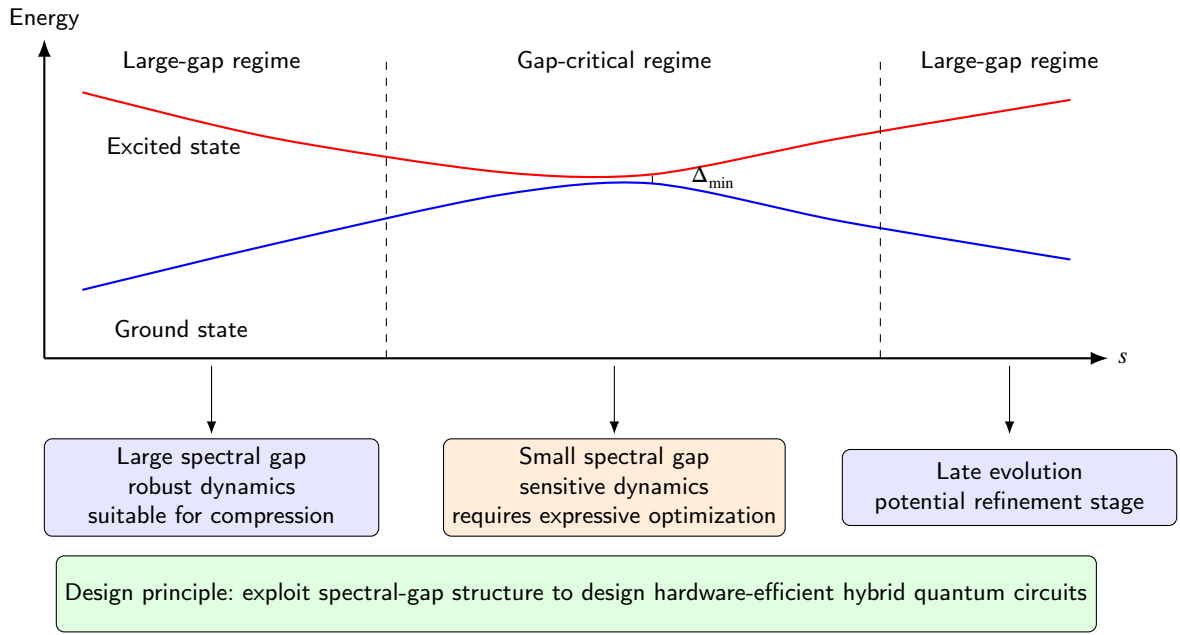
\begin{figure*}[t]
\centering
\begin{tikzpicture}[
>=Latex,
box/.style={draw, rounded corners=3pt, minimum width=4.4cm, minimum height=1cm, align=center, font=\small},
design/.style={draw, rounded corners=3pt, minimum width=12cm, minimum height=1cm, align=center, fill=green!12, font=\small}
]

\draw[thick,->] (0,0) -- (14,0) node[right] {$s$};
\draw[thick,->] (0,0) -- (0,4.2) node[above] {Energy};

\draw[thick,blue]
plot[smooth] coordinates
{(0.5,0.9) (3,1.5) (6,2.15) (8,2.30) (10.5,1.8) (13.5,1.3)};

\draw[thick,red]
plot[smooth] coordinates
{(0.5,3.5) (3,2.9) (6,2.45) (8,2.42) (10.5,2.9) (13.5,3.4)};

\node at (1.7,2.8) {Excited state};
\node at (1.8,0.4) {Ground state};

\draw[dashed] (8,2.30) -- (8,2.42);
\node at (8.8,2.38) {$\Delta_{\min}$};

\draw[dashed] (4.5,0) -- (4.5,4);
\draw[dashed] (11,0) -- (11,4);

\node at (2.2,3.9) {Large-gap regime};
\node at (7.5,3.9) {Gap-critical regime};
\node at (12.7,3.9) {Large-gap regime};

\draw[->] (2.2,-0.1) -- (2.2,-1.0);
\draw[->] (7.5,-0.1) -- (7.5,-1.0);
\draw[->] (12.7,-0.1) -- (12.7,-1.0);

\node[box, fill=blue!10] at (2.2,-1.7)
{Large spectral gap\\robust dynamics\\suitable for compression};

\node[box, fill=orange!15] at (7.5,-1.7)
{Small spectral gap\\sensitive dynamics\\requires expressive optimization};

\node[box, fill=blue!10] at (12.7,-1.7)
{Late evolution\\potential refinement stage};

\node[design] at (7,-3.1)
{Design principle: exploit spectral-gap structure to design hardware-efficient hybrid quantum circuits};

\end{tikzpicture}

\caption{
Spectral-gap-aware intuition for hybrid circuit design.
The adiabatic evolution path exhibits regions with different spectral-gap structures.
Large-gap regimes are generally more robust and can tolerate circuit compression,
whereas the gap-critical region near the minimum spectral gap $\Delta_{\min}$
is more sensitive and motivates retaining expressive variational circuit layers.
}
\label{fig:spectral_gap_intuition}
\end{figure*}

\section{Background and Related Work}

To motivate the proposed framework, we review related work from an application-oriented perspective, covering transportation optimization models, quantum algorithmic approaches for combinatorial search, and hardware-aware methods for making such algorithms executable on near-term devices.

\subsection{Transportation Optimization Problems}

Transportation systems involve numerous planning and operational problems that can be formulated as combinatorial optimization tasks. Among the most widely studied are the Vehicle Routing Problem (VRP) \citep{dantzig1959truck}, the Traveling Salesman Problem (TSP) \citep{matai2010traveling}, and the Facility Location Problem (FLP) \citep{cornuejols1983uncapicitated}. These models form the foundation of many logistics and supply chain planning applications.

The VRP determines optimal routes for a fleet of vehicles serving geographically distributed customers while minimizing travel distance or cost. Numerous variants have been proposed to address practical considerations such as vehicle capacities, service time windows, and stochastic travel times \citep{laporte2009fifty}. Similarly, facility location models determine the optimal placement of infrastructure such as warehouses, distribution centers, or charging stations in order to balance facility construction costs with transportation costs \citep{farahani2009facility}.

Classical solution methods include exact approaches based on mixed-integer programming as well as heuristic and metaheuristic algorithms. While exact algorithms can solve moderate-sized instances, complexity of these problems grows rapidly with the number of customers or facilities, motivating research into alternative computational paradigms. For transportation contexts, the critical challenge is not merely NP-hardness but the combination of large feasible spaces and constraint structures that make naive sampling highly infeasible.

\subsection{Quantum Algorithms for Combinatorial Optimization}

Two major classes of quantum optimization algorithms have been proposed: adiabatic quantum computing \citep{farhi2000adiabatic} and variational quantum algorithms \citep{mcclean2016vqe}. Adiabatic quantum computing solves optimization problems by evolving a quantum system from a simple driver Hamiltonian to a problem Hamiltonian whose ground state encodes the optimal solution. Quantum annealers implement analog time-dependent evolution for Ising/QUBO objectives and have enabled many of the earliest logistics and routing demonstrations \citep{feld2019hybrid, borowski2020new}. Similarly, large instances of the facility location problem have also been investigated using a D-Wave quantum annealer \citep{li2026quantum}. On gate-based quantum devices, this evolution must be approximated through Trotterized time evolution, which decomposes continuous dynamics into discrete gate operations. However, this approach often leads to deep circuits containing large numbers of two-qubit gates, making it difficult to implement on current hardware.

Variational quantum algorithms attempt to address these limitations by using parameterized quantum circuits whose parameters are optimized using classical optimization routines. The quantum Approximate Optimization Algorithm (QAOA) is one of the most widely studied approaches in this class. QAOA constructs a parameterized circuit consisting of alternating applications of a problem Hamiltonian and a mixer Hamiltonian \citep{farhi2014qaoa}. While QAOA can significantly reduce circuit depth relative to direct Trotterization of adiabatic evolution, its performance depends heavily on the number of layers and the success of classical parameter optimization.

QUBO/Ising mappings are widely used to provide a common interface between transportation optimization models and quantum optimization primitives \citep{lucas2014ising}. Constraints (e.g., one-hot assignments, flow conservation, capacity) are typically embedded through penalty terms, creating energy landscapes with many infeasible states. Constrained-QAOA and related ``alternating operator'' frameworks construct mixers that preserve feasible subspaces \citep{hadfield2019quantum,wang2020xy}, but hardware realizations of such mixers can be demanding, motivating connectivity-aware approximations.

\subsection{Hardware-Efficient Quantum Algorithms and Circuit Compression}

Given NISQ limits of circuit depth \citep{preskill2018quantum}, compilation and circuit synthesis are central to practical studies. Importantly, the connectivity structure of the target quantum hardware can introduce additional constraints that affect circuit depth. In an effort to avoid this, linear chain QAOA (LC-QAOA) has been recently proposed \citep{wang2025depth}, which entangles only neighboring qubits. Other approaches which acknowledge that only low depth QAOA can be run practically try to improve performance by using an initial state corresponding to the solution of a relaxed convex problem in a method known as Warm-starting \citep{egger2021warm}. Yet other methods such as recursive-QAOA \citep{bravyi2022hybrid} iteratively reduce the size of the problem by identifying correlated clusters of variables, thus requiring smaller circuits. 

Approximate quantum compilation replaces a target unitary (or the action of a circuit on a state) by a shorter circuit optimized classically while maintaining fidelity. Tensor networks have been applied to many aspects of quantum computing including quantum simulation, circuit synthesis, and error correction \citep{berezutskii2025tensor}. Recent work has explored the use of tensor-network representations (e.g., MPS) to efficiently optimize such approximations for portions of time-evolution circuits, enabling efficient classical optimization over circuit parameters while preserving the essential dynamics of the original evolution \citep{robertson2025approximate}. The method has been shown to reduce depth by an order of magnitude as compared to other matrix-product-state circuit synthesis algorithms, yet has not been applied towards combinatorial optimization.

\subsection{Quantum Computing in Transportation: Application-led Benchmarking}

In transportation and logistics, prior quantum optimization work has concentrated on quantum annealing-based approaches to routing \citep{weinberg2023supply} and facility location \citep{li2026quantum}. From a gate-based perspective, previous works have studied facility location through QUBO formulations and QAOA, but their results were reported in simulation rather than on quantum hardware \citep{moncayo2025quantum}. Similarly, the application of QAOA for the TSP has been examined in simulation, comparing mixers and parameter training schedules \citep{qian2023comparative}. The heterogeneous VRP has also been analyzed for circuit depth and the discrimination of good and bad solutions by QAOA with simulated results \citep{fitzek2024applying}. 

A recent review of quantum computing in transportation engineering concluded that in the NISQ era priority should be given to robustness and scalability \citep{somvanshi2026quantum}. Meanwhile, application-focused studies in transportation suggest that near-term quantum computing is better viewed through the lens of application-led benchmarking and integration with classical optimization workflows, rather than through claims of standalone quantum advantage \citep{bentley2022quantum,du2026overcoming}. 
In this perspective, quantum algorithms are viewed as components within hybrid computational pipelines that may support tasks such as candidate solution generation, warm-start initialization, or problem decomposition. Our work follows this application-driven approach by focusing on reproducible tradeoffs between circuit depth and solution feasibility, and by examining how specific circuit design choices translate into practical performance for transportation decision-support problems.

Despite growing interest in applying quantum optimization methods to logistics and transportation applications \citep{zhuang2024quantum}, two important challenges remain. First, realistic transportation optimization problems frequently require circuit depths that exceed the capabilities of current quantum hardware \citep{azfar2025quantum}. Second, the structure of adiabatic quantum evolution -- particularly the existence of distinct large-gap regimes near the beginning and end of the evolution and gap-critical regions in the middle -- has not yet been fully exploited in the design of hybrid quantum optimization algorithms. While tensor networks have been proposed to solve quantum optimization \citep{binimelis2024quantum}, and time evolution circuits are known to compress efficiently compared to random circuits \citep{guo2025efficient}, approximate quantum compilation has not yet been used for transportation optimization problems. 

The present work addresses these challenges by systematically studying prefix compression of Trotterized annealing circuits together with hybrid architectures that combine compressed annealing segments with variational optimization layers. By evaluating these methods on transportation optimization problems, we aim to provide insight into how hardware-aware circuit design can improve the practical feasibility of quantum optimization methods for real-world logistics and transportation planning applications. It is effectively a cross-problem, hardware-grounded analysis of how compiled prefix initialization, ansatz choice, and depth/noise tradeoffs interact across FLP, VRP, and TSP.

\section{Transportation Problem Models and Binary Encodings}
\label{sec:models}

We evaluate three canonical transportation optimization models (TSP, VRP, FLP). Each is encoded as a binary optimization problem and then mapped to a diagonal cost Hamiltonian for quantum optimization experiments.

\subsection{Traveling Salesman Problem}

A traveling salesman problem (TSP) is defined on a set of \(n\) cities,
\[
\mathcal{V} = \{1,2,\dots,n\},
\]
with pairwise travel costs \(d_{ij}\) representing the distance (or more generally, the cost) of traveling from city \(i\) to city \(j\).

To encode the tour, we introduce binary decision variables
\begin{equation}
x_{r,c} \in \{0,1\},
\qquad r,c \in \{1,\dots,n\},
\end{equation}
where \(x_{r,c}=1\) indicates that city \(c\) is visited at position \(r\) in the tour, and \(x_{r,c}=0\) otherwise. This formulation uses a permutation-based encoding in which the tour is represented as an \(n \times n\) binary assignment matrix.

A valid TSP tour must satisfy two sets of one-hot constraints. First, each tour position must be occupied by exactly one city. These \emph{row constraints} are given by
\begin{equation}
\sum_{c=1}^{n} x_{r,c} = 1,
\qquad \forall r \in \{1,\dots,n\}.
\label{eq:tsp_row_constraint}
\end{equation}

Second, each city must appear exactly once in the tour. These \emph{column constraints} are written as
\begin{equation}
\sum_{r=1}^{n} x_{r,c} = 1,
\qquad \forall c \in \{1,\dots,n\}.
\label{eq:tsp_column_constraint}
\end{equation}

Together, \eqref{eq:tsp_row_constraint} and \eqref{eq:tsp_column_constraint} ensure that the binary matrix \(X=[x_{r,c}]\) corresponds to a permutation matrix, and therefore represents a Hamiltonian cycle visiting each city exactly once.

The total tour cost is obtained by summing the travel cost between consecutive tour positions. Letting the tour wrap around cyclically so that position \(n+1\) is identified with position \(1\), the objective function is
\begin{equation}
C(X)
=
\sum_{r=1}^{n}
\sum_{c_1=1}^{n}
\sum_{c_2=1}^{n}
d_{c_1 c_2}\,
x_{r,c_1}\,x_{r+1,c_2},
\label{eq:tsp_cost}
\end{equation}
where the index \(r+1\) is understood modulo \(n\). Equivalently, one may write \(x_{n+1,c} \equiv x_{1,c}\) for all \(c\).

The TSP can therefore be formulated as the following binary optimization problem:
\begin{align}
\min_{x} \quad
&
\sum_{r=1}^{n}
\sum_{c_1=1}^{n}
\sum_{c_2=1}^{n}
d_{c_1 c_2}\,
x_{r,c_1}\,x_{r+1,c_2}
\label{eq:tsp_full_obj}
\\
\text{s.t.} \quad
&
\sum_{c=1}^{n} x_{r,c} = 1,
\qquad \forall r \in \{1,\dots,n\},
\label{eq:tsp_full_row}
\\
&
\sum_{r=1}^{n} x_{r,c} = 1,
\qquad \forall c \in \{1,\dots,n\},
\label{eq:tsp_full_col}
\\
&
x_{r,c} \in \{0,1\},
\qquad \forall r,c \in \{1,\dots,n\}.
\label{eq:tsp_full_binary}
\end{align}

This permutation-based representation is widely used in quadratic and quantum optimization formulations of the TSP because it expresses feasibility through structured one-hot constraints and captures the tour length through pairwise interactions between adjacent positions in the route \citep{lucas2014ising}.

\subsection{Vehicle Routing Problem}

We consider a capacitated vehicle routing problem (VRP) defined on a directed graph
\[
G = (\mathcal{N}, \mathcal{E}),
\]
where \(\mathcal{N} = \{0,1,2,\dots,n\}\) denotes the set of nodes and \(\mathcal{E}\) the set of directed edges between nodes. Node \(0\) represents the depot, while nodes \(1,\dots,n\) correspond to customer locations. Let \(w_{ij}\) denote the travel cost (or distance) associated with traversing the directed edge from node \(i\) to node \(j\).

The goal of the VRP is to determine a set of \(K\) vehicle routes that start and end at the depot while visiting each customer exactly once, such that the total travel cost is minimized. We assume that customer demand is satisfied upon visitation and that no time-window constraints are imposed.

To encode the routing decisions, we introduce binary variables
\begin{equation}
x_{ij} \in \{0,1\}, \qquad i,j \in \mathcal{N},
\end{equation}
where \(x_{ij}=1\) indicates that a vehicle travels directly from node \(i\) to node \(j\), and \(x_{ij}=0\) otherwise.

The total travel cost of the routing solution is given by
\begin{equation}
\min_{x}
\sum_{i \in \mathcal{N}} \sum_{j \in \mathcal{N}} w_{ij} x_{ij}.
\label{eq:vrp_objective}
\end{equation}

Feasible routes must satisfy several structural constraints.

First, each customer must be visited exactly once. This is enforced by requiring exactly one outgoing edge from every customer node:
\begin{equation}
\sum_{j \in \mathcal{N}\setminus\{0\}} x_{ij} = 1,
\qquad \forall i \in \mathcal{N}\setminus\{0\}.
\label{eq:vrp_visit_once}
\end{equation}

Second, route continuity must be maintained. If a vehicle arrives at a node, it must also depart from that node. This flow conservation condition can be expressed as
\begin{equation}
\sum_{j \in \mathcal{N}} x_{ij}
=
\sum_{j \in \mathcal{N}} x_{ji},
\qquad \forall i \in \mathcal{N}.
\label{eq:vrp_flow}
\end{equation}

Third, the number of vehicles departing from and returning to the depot must equal the fleet size \(K\):
\begin{align}
\sum_{j \in \mathcal{N}} x_{0j} &= K, \label{eq:vrp_depot_out}\\
\sum_{i \in \mathcal{N}} x_{i0} &= K. \label{eq:vrp_depot_in}
\end{align}

Finally, additional constraints are required to eliminate disconnected cycles that do not include the depot. These \emph{subtour elimination constraints} ensure that every subset of customer nodes connects to the rest of the network. For any subset \(S \subset \mathcal{N}\) with \(2 \le |S| \le n-1\), the following condition must hold:
\begin{equation}
\sum_{i \in S} \sum_{j \notin S} x_{ij} \ge 1.
\label{eq:vrp_subtour}
\end{equation}

Constraint \eqref{eq:vrp_subtour} guarantees that at least one edge leaves any proper subset of nodes, thereby preventing isolated cycles that do not connect to the depot.

Without these subtour elimination constraints, solutions satisfying the other conditions may still contain disconnected loops. For example, in an eight-node instance with two vehicles, the routes
\[
0 \rightarrow 1 \rightarrow 2 \rightarrow 0, \qquad
0 \rightarrow 3 \rightarrow 4 \rightarrow 0,
\]
could coexist with a disconnected cycle
\[
5 \rightarrow 6 \rightarrow 7 \rightarrow 5,
\]
which satisfies the assignment and flow constraints but does not constitute a valid routing solution. The subtour elimination constraints prevent such configurations.

Combining the objective function \eqref{eq:vrp_objective} with constraints
\eqref{eq:vrp_visit_once}–\eqref{eq:vrp_subtour} yields a standard edge-based mixed-integer formulation of the vehicle routing problem. As shown in \citep{azfar2025quantum}, this formulation requires the fewest quantum resources for VRP instances with $n \leq 7$, although this advantage does not persist asymptotically.

\subsection{Facility Location Problem}

An uncapacitated facility location problem consists of a set of customers
\[
\mathcal{I}=\{1,2,\dots,N\}
\]
and a set of candidate facilities
\[
\mathcal{J}=\{1,2,\dots,M\}.
\]

For each facility \(j \in \mathcal{J}\), let \(f_j\) denote the fixed setup cost of opening facility \(j\). For each customer \(i \in \mathcal{I}\) and facility \(j \in \mathcal{J}\), let \(c_{ij}\) denote the service cost of assigning customer \(i\) to facility \(j\).

We define the following binary decision variables:
\[
y_j =
\begin{cases}
1, & \text{if facility } j \text{ is opened},\\
0, & \text{otherwise},
\end{cases}
\qquad \forall j \in \mathcal{J},
\]
\[
x_{ij} =
\begin{cases}
1, & \text{if customer } i \text{ is assigned to facility } j,\\
0, & \text{otherwise},
\end{cases}
\]
\[
\qquad \forall i \in \mathcal{I},\; j \in \mathcal{J}.
\]

The objective is to minimize the total facility opening cost and customer assignment cost:
\begin{equation}
\min \; \sum_{j \in \mathcal{J}} f_j y_j
\;+\;
\sum_{i \in \mathcal{I}} \sum_{j \in \mathcal{J}} c_{ij} x_{ij}.
\label{eq:flp_objective}
\end{equation}

This optimization is subject to two sets of constraints.

First, each customer must be assigned to exactly one facility:
\begin{equation}
\sum_{j \in \mathcal{J}} x_{ij} = 1,
\qquad \forall i \in \mathcal{I}.
\label{eq:flp_assignment}
\end{equation}

Second, a customer can only be assigned to a facility if that facility is open:
\begin{equation}
x_{ij} \le y_j,
\qquad \forall i \in \mathcal{I},\; j \in \mathcal{J}.
\label{eq:flp_open}
\end{equation}

Finally, all variables are binary:
\begin{equation}
x_{ij} \in \{0,1\},
\qquad
y_j \in \{0,1\},
\qquad \forall i \in \mathcal{I},\; j \in \mathcal{J}.
\label{eq:flp_binary}
\end{equation}

Combining \eqref{eq:flp_objective}--\eqref{eq:flp_binary}, the uncapacitated facility location problem can be written compactly as
\begin{align}
\min_{x,y} \quad
& \sum_{j \in \mathcal{J}} f_j y_j
+ \sum_{i \in \mathcal{I}} \sum_{j \in \mathcal{J}} c_{ij} x_{ij}
\\
\text{s.t.} \quad
& \sum_{j \in \mathcal{J}} x_{ij} = 1,
&& \forall i \in \mathcal{I},
\\
& x_{ij} \le y_j,
&& \forall i \in \mathcal{I},\; j \in \mathcal{J},
\\
& x_{ij} \in \{0,1\},
&& \forall i \in \mathcal{I},\; j \in \mathcal{J},
\\
& y_j \in \{0,1\},
&& \forall j \in \mathcal{J}.
\end{align}

\subsection{From Binary Models to QUBO/Ising Hamiltonians}

Each problem is mapped to a QUBO objective (and penalties for constraints) and then to an Ising-type Hamiltonian \(H_C\) diagonal in the computational basis \citep{lucas2014ising}. We write the problem Hamiltonian in Pauli-\(Z\) form:
\[
H_C = \sum_i h_i Z_i + \sum_{i<j} J_{ij} Z_iZ_j,
\]
with an affine mapping between decision bits \(b_i\in\{0,1\}\) and spin variables (eigenvalues of \(Z_i\)). Penalty weights are selected to separate feasible and infeasible energies:
\[
C_{\mathrm{QUBO}}(b)= C_{\mathrm{obj}}(b) + \lambda\,C_{\mathrm{pen}}(b),
\]
As a rule of thumb, $\lambda > \sum_{i\neq j} |w_{i,j}|$ is considered sufficient \citep{harwood2021formulating}. Prior empirical results \citep{azfar2025quantum} suggest that using a factor of two relative to this bound can improve feasibility ratios during optimization. This heuristic ties $\lambda$ to the native cost scale of the problem, ensuring that violating a constraint is always more expensive than any reduction in travel cost. In practice, $\lambda$ should be large enough to dominate objective improvements while avoiding numerical conditioning issues during transpilation and parameter optimization. Therefore, a global penalty is set as
\[
\lambda = 2 \sum_{i \neq j} |w_{ij}|.
\]
After conversion to QUBO form, all coefficients are normalized by dividing by the magnitude of the largest coefficient. This scaling preserves the $\arg\min$ of the objective while improving numerical conditioning for parameter optimization and circuit transpilation. In addition, normalization keeps the effective Hamiltonian coefficients within a bounded range, which improves stability of variational parameter updates and facilitates hardware-compatible gate synthesis.

\section{Hybrid Quantum Optimization Framework}
\label{sec:framework}

This section defines the circuit families evaluated and the diagnostics used to interpret performance. The goal is reproducibility and transportation-relevant interpretability, not algorithmic novelty.

\subsection{Digitized Adiabatic Evolution}

We consider an interpolation Hamiltonian
\begin{equation}
H(s) = (1-s)H_B + sH_C,
\end{equation}
where \(H_B=\sum_i X_i\) is a transverse-field driver and \(H_C\) is the diagonal cost Hamiltonian. 

With \(N_{\mathrm{steps}}\) Trotter steps and schedule \(s_k\in[0,1]\), the digitized annealing circuit is \begin{equation} U_{\mathrm{anneal}} \approx \prod_{k=0}^{N_{\mathrm{steps}}-1} \exp\!\bigl(-i\Delta t\,H(s_k)\bigr), \qquad \Delta t = T/N_{\mathrm{steps}}. \label{eq:anneal_trotter} \end{equation} 
where \(\Delta t=1/N_{\mathrm{steps}}\) and \(s_k\) is obtained from a linear schedule. Each trotter step is implemented in qiskit via a \texttt{PauliEvolutionGate} over the full qubit register, and the initial state is \(|+\rangle^{\otimes n}\).

In the experiments, total annealing time of \(T=1\) is used after coefficient normalization, so that the effective dynamics are governed by the relative scales of the driver and problem Hamiltonians rather than by an arbitrary global phase factor.

\subsection{AQC-Tensor Prefix Compression}
\label{subsec:aqc}

To reduce circuit depth while preserving the evolution, approximate quantum compilation (AQC) is used to compress prefixes of the Trotterized annealing circuit via tensor-network methods. For \(m=1,\ldots,6\), the target prefix is
\[
U_{\mathrm{target}}^{(m)} = \prod_{k=0}^{m-1} e^{-i\Delta t H(s_k)},
\]
and the goal is to find a shallower circuit \(C^{(m)}\) that approximates this evolution on the initial state with fidelity
\[
F\bigl(C^{(m)}|\psi_{\mathrm{init}}\rangle,\; U_{\mathrm{target}}^{(m)}|\psi_{\mathrm{init}}\rangle\bigr) \ge \eta .
\]

The compression uses the AQC-Tensor workflow for time-evolution circuits \citep{qiskit-addon-aqc-tensor}. A parameterized shallow ans\"atz is constructed from a reduced ``good prefix'', and its parameters are optimized so that its output state matches that of the target prefix. During optimization, both the target prefix and the shallow ans\"atz are simulated using a tensor-network (MPS) simulator, and the ans\"atz parameters are updated to maximize the fidelity between the two states until the threshold \(\eta\) or a maximum iteration limit is reached.

In practice, the target prefix is generated at twice the resolution of the layers it replaces, i.e.,
\[
\Delta t_{\text{target}} = \Delta t / 2 .
\]
The shallow ans\"atz is constructed as a parameterized version of a good prefix consisting of \(\lceil m/3 \rceil\) Trotter steps at this scaled resolution. Thus, the higher resolution Trotterized annealing evolution is approximated by a circuit implementing the same evolution with substantially fewer Trotter steps.

The resulting compressed prefix \(C^{(m)}\) is concatenated with the remaining uncompressed evolution,
\[
U^{(m)} = C^{(m)} \prod_{k=m}^{N_{\mathrm{steps}}-1} e^{-i\Delta t H(s_k)}.
\]


\subsection{Hybrid Replacement with Standard QAOA}

To explore variational alternatives to the remaining uncompressed Trotter steps, we replace the trotterized tail of each compressed circuit \(U^{(m)}\) with a layer of standard QAOA. For a circuit with \(N_{\mathrm{steps}}=10\) and \(m\)-step compressed prefix, we apply an ans\"atz of the form
\[
U_{\mathrm{QAOA}}^{(m,p)} = C^{(m)} \prod_{\ell=1}^{p} e^{-i\beta_\ell H_M} e^{-i\gamma_\ell H_P},
\]
where \(p\) is the number of QAOA layers, \(H_M=\sum_i X_i\) is the mixer Hamiltonian, and \(H_P=H_{\mathrm{problem}}\). We test several values of \(p\) (e.g., \(p\in\{1 \ ... \ 6\}\)) to evaluate the tradeoff between expressivity, depth, and solution quality.

For each \((m,p)\) configuration, we optimize the QAOA angles \(\{\beta_\ell,\gamma_\ell\}\) using classical optimizers and report the percentage of feasible bitstrings and objective value distributions.

\subsection{Hybrid Replacement with Linear Chain QAOA}

To reduce two-qubit gate depth further, we consider a structured variational ans\"atz inspired by a linear chain (LC-QAOA) \citep{wang2025depth}. Long range qubit connections require SWAP gates, so the LC-QAOA ans\"atz modifies the cost Hamiltonian to restrict interactions along a linear topology, reducing the required entangling gates. We construct circuits
\[
U_{\mathrm{LC}}^{(m,p)} = C^{(m)} \prod_{\ell=1}^{p} e^{-i\beta_\ell H_{M}} e^{-i\gamma_\ell H_{P,\mathrm{LC}}},
\]
where \(H_{P,\mathrm{LC}}\) contains only nearest-neighbor \(ZZ\) couplings. We evaluate the same set of layer counts \(p\) as in the standard QAOA case and perform identical optimization and sampling procedures.

\subsection{Optimization with CVaR}

For diagonal cost Hamiltonians, the objective value can be estimated directly from measurement samples. All QAOA parameter optimization in this study uses the Conditional Value-at-Risk (CVaR) objective \citep{barkoutsos2020cvar}, which has been shown to perform well on noisy quantum devices. Instead of optimizing the mean cost over all samples, CVaR focuses on the lowest-cost portion of the measurement distribution, which makes the optimization less sensitive to noise and sampling fluctuations.

For a given parameter setting, measurement outcomes are evaluated under the classical cost function and sorted by cost. The CVaR objective optimizes the average cost of the best \(\alpha\) fraction of samples,
\[
\mathrm{CVaR}_\alpha(C) = \mathbb{E}[C \mid C \le q_\alpha],
\]
where \(q_\alpha\) denotes the empirical \(\alpha\)-quantile of the sampled costs.

In practice, the aggregation parameter \(\alpha\) is chosen adaptively based on the estimated circuit fidelity. Specifically, we extract the layer fidelity `$\mathrm{lf}$' for an $n$-qubit circuit from backend calibration data and compute the error per layered gate (EPLG) as
\begin{equation}
\mathrm{EPLG} = 1 - \mathrm{lf}^{\frac{1}{n-1}}.
\end{equation}
The corresponding two-qubit gate fidelity is $\mathrm{fid}_{\mathrm{cx}} = 1 - \mathrm{EPLG}$, from which we define a noise amplification factor scaled by two-qubit gate depth $D_{2q}$ as
\begin{equation}
\gamma_{\mathrm{circ}} = \frac{1}{\mathrm{fid}_{\mathrm{cx}}^2}  \cdot D_{2q}
\end{equation}
Finally, the CVaR parameter is set as
\begin{equation}
\alpha = \frac{1}{\sqrt{\gamma_{\mathrm{circ}}}}.
\end{equation}
 This heuristic concentrates the optimization on the fraction of samples that are most likely to correspond to correct circuit executions while discarding high-cost outcomes introduced by hardware noise.

\subsection{Sampling-accessible Diagnostics: Variance and Susceptibility}

Direct measurement of the instantaneous spectral gap is generally infeasible on current NISQ devices because it requires full spectral information or coherent phase estimation. Instead, gap-sensitive diagnostics are estimated from measurement statistics produced by the quantum circuit. Two quantities are considered: the energy variance of the cost Hamiltonian and the perturbative susceptibility of an observable.

\paragraph{Energy variance.}

Let $|\psi\rangle$ denote the state prepared by the circuit and $H_C$ the diagonal cost Hamiltonian. The energy variance is
\[
\mathrm{Var}(H_C) = \langle H_C^2 \rangle - \langle H_C \rangle^2 .
\]

For QUBO or Ising problems, $H_C$ is diagonal in the computational basis, so the moments of the cost distribution can be estimated directly from sampled bitstrings $z$ with probability $p(z)$ and cost $C(z)$:
\[
\langle H_C \rangle \approx \sum_z p(z) C(z), 
\qquad
\langle H_C^2 \rangle \approx \sum_z p(z) C(z)^2 .
\]

If the prepared state is close to an eigenstate, the variance is small. Increased variance indicates mixing across multiple energy levels, which often occurs near narrow spectral gaps during annealing-like evolution.

\paragraph{Perturbative susceptibility.}

A complementary diagnostic measures the response of the system to a small Hamiltonian perturbation. At interpolation point $s_m$,
\[
H(s_m) = (1-s_m)H_B + s_m H_C .
\]

A perturbed Hamiltonian
\[
H_\pm = H(s_m) \pm \lambda V
\]
is applied for a short probe evolution, and the expectation value of $V$ is measured. The susceptibility is estimated by a finite-difference approximation
\[
\chi_V \approx 
\frac{\langle V \rangle_{+\lambda} - \langle V \rangle_{-\lambda}}{2\lambda}.
\]

In this work the observable is chosen as the driver Hamiltonian
\[
V = H_B = \sum_i X_i ,
\]
which probes how strongly the state responds to transverse-field perturbations.

\paragraph{Interpretation.}

Large values of the energy variance or susceptibility indicate regions where the quantum state becomes highly sensitive to small Hamiltonian changes. Such behavior is commonly associated with avoided crossings or narrow spectral gaps, providing experimentally accessible indicators of gap-sensitive regions in the optimization landscape.

\section{Experimental Design}
\label{sec:exp}

The experimental workflow consists of four stages: (i) instance generation and QUBO formulation, (ii) circuit construction and compilation, (iii) parameter optimization using CVaR-based objectives, and (iv) sampling-based evaluation of feasibility and solution quality.
All circuits are executed on IBM quantum hardware using the Qiskit Runtime primitives, with sampling-based evaluation performed under realistic device noise and connectivity constraints.

\begin{algorithm}[!htbp]
\caption{Experimental pipeline for AQC prefix compression and (LC-)QAOA substitution}
\label{alg:pipeline}
\begin{algorithmic}[1]
\Require Problem Hamiltonian $H_{\mathrm{prob}}$, driver $H_{\mathrm{init}}$, schedule $\{s_k\}_{k=0}^{9}$, steps $N=10$, prefix set $\mathcal{A}=\{1,\dots,6\}$, QAOA depths $\mathcal{P}=\{1,\dots,6\}$.
\For{$\alpha \in \mathcal{A}$}
  \State Build target prefix $U_{\mathrm{pref}}^{(\alpha)}$ from first $\alpha$ Trotter steps
  \State Compress via AQC $\rightarrow \widetilde U_{\mathrm{pref}}^{(\alpha)}$ and record $F_\alpha$
  \State Build $U_{\mathrm{AQC\text{-}trot}}^{(\alpha)} = U_{\mathrm{tail}}^{(\alpha)}\widetilde U_{\mathrm{pref}}^{(\alpha)}$
  \State Evaluate $\Pfeas$, $\Popt$, $\Efeas$, and $\Dtwo$
  \For{$p \in \mathcal{P}$}
    \State Build $U_{\mathrm{AQC\text{-}QAOA}}^{(\alpha,p)} = U_{\mathrm{QAOA}}(p)\widetilde U_{\mathrm{pref}}^{(\alpha)}$
    \State Optimize and record best metrics
    \State Build $U_{\mathrm{AQC\text{-}LC}}^{(\alpha,p)} = U_{\mathrm{LC\text{-}QAOA}}(p)\widetilde U_{\mathrm{pref}}^{(\alpha)}$ 
    \State Optimize and record metrics as above
  \EndFor
\EndFor
\end{algorithmic}
\end{algorithm}

\subsection{Problem Instances and Transportation Interpretation}

The experimental evaluation considers optimization instances representing three levels of transportation decision-making: single-tour routing, multi-vehicle routing, and strategic facility placement. These correspond respectively to the Traveling Salesperson Problem (TSP), the Vehicle Routing Problem (VRP), and the Facility Location Problem (FLP), which serve as canonical abstractions for operational, tactical, and strategic planning tasks in transportation systems.

Instance sizes are selected to balance two considerations. First, the problems must contain sufficient combinatorial structure to produce nontrivial feasibility constraints and cost tradeoffs typical of real transportation models. Second, they must remain within the limits of classical quantum tensor-network simulation, which are required to run the AQC functions. For each instance, the optimal classical solution is computed using exact solvers or exhaustive search, allowing solution quality to be measured relative to a known optimum.

Table~\ref{tab:instance_summary} summarizes the key characteristics of each instance, including the problem type, instance size, number of binary decision variables (equivalently qubits), the objective, and constraints used in the QUBO formulation.

While quantum optimization performance is often reported using energy-based and sampling-based metrics, transportation decision-making requires evaluating the quality and usability of candidate solutions under practical constraints. In this work, we interpret quantum sampling outcomes as a candidate-generation process for transportation plans, where each measured bitstring corresponds to a routing, assignment, or siting configuration.

Under this interpretation, the quantum algorithm is not viewed as directly producing a single optimal solution, but rather as generating a distribution over candidate solutions from which a planner can select. This aligns with common practice in transportation analytics, where heuristic methods generate multiple feasible alternatives that are subsequently evaluated or refined.

\begin{table*}[!htb]
\centering
\caption{Summary of optimization instances used in the experiments. All instances are small-scale to enable exact evaluation of solution quality and detailed analysis of quantum circuit behavior.}
\label{tab:instance_summary}
\begin{tabular}{llllll}
\toprule
Problem & Instance size  & Qubits & Objective     & Constraints  & Notes                                                                             \\ 
\midrule
FLP     & \begin{tabular}[c]{@{}l@{}}5 customers,\\ 2 facilities\end{tabular} & 12     & \begin{tabular}[c]{@{}l@{}}Facility cost, \\ assignment cost\end{tabular} & \begin{tabular}[c]{@{}l@{}}Assignment,\\ open-if-assigned\end{tabular}                     & \begin{tabular}[c]{@{}l@{}}Grid-based\\ placement\end{tabular}                    \\
VRP     & \begin{tabular}[c]{@{}l@{}}4 stops,\\ 2 vehicles\end{tabular}       & 14     & Total travel cost                                                              & \begin{tabular}[c]{@{}l@{}}In/out degree,\\ depot flow,\\ subtour elimination\end{tabular} & \begin{tabular}[c]{@{}l@{}}Edge-based\\ formulation\end{tabular}                  \\
TSP     & 5 cities                                                            & 16     & Total tour length                                                              & \begin{tabular}[c]{@{}l@{}}Reduced row/column\\ one-hot constraints\end{tabular}           & \begin{tabular}[c]{@{}l@{}}Assignment-based \\ encoding, fixed depot\end{tabular} \\ 
\bottomrule
\end{tabular}
\end{table*}

\subsection{Evaluation Metrics and Performance Criteria}

Evaluating quantum optimization algorithms in transportation settings requires bridging two perspectives: (i) quantum circuit performance under hardware constraints, and (ii) the quality and usability of generated solutions for decision-making. Rather than relying solely on expectation values of the cost Hamiltonian, which may not directly correspond to actionable plans, we evaluate circuits as \emph{candidate generators} that produce distributions over feasible solutions. These metrics are evaluated using measurement outcomes obtained from hardware execution, reflecting the stochastic and noisy nature of near-term quantum devices.



\paragraph{Hardware efficiency.}
We measure the transpiled \textbf{two-qubit gate depth} ($D_{2q}$) as a proxy for circuit feasibility on near-term devices. Since two-qubit operations dominate noise and execution time, this metric captures the practical hardware cost of each configuration.

\paragraph{Optimization effort.}
We report the number of \textbf{classical optimizer iterations} required for convergence of the variational parameters. This reflects the difficulty of navigating the parameter landscape as circuit depth and expressivity increase.

\paragraph{Feasible solution generation.}
Given a fixed sampling budget of 10,000 samples, we evaluate the \textbf{feasible sample count}: the number of measurement outcomes satisfying all problem constraints, interpreted as the probability of obtaining a usable transportation plan from a single circuit execution.

\paragraph{Solution diversity.}
We measure the number of \textbf{unique feasible solutions} observed in the sampled distribution. This captures the ability of the algorithm to generate multiple distinct candidate plans, which is critical in transportation settings where alternative routing or assignment options are valuable.

\paragraph{Solution quality.}
We report the \textbf{average objective value} across samples, which reflect the quality of candidate solutions produced under realistic sampling constraints.

\paragraph{Diagnostic metrics.}
In addition to the above evaluation metrics, we use variance of the problem Hamiltonian and perturbative susceptibility as diagnostic tools to analyze the structure of compressed annealing prefixes. These quantities are not used for performance comparison, but rather to guide understanding of where compression may affect the underlying optimization dynamics.

These metrics collectively characterize each circuit as a stochastic generator of candidate transportation plans under hardware constraints. In contrast to just expectation-based evaluations, this perspective emphasizes the tradeoff between circuit depth, feasibility, diversity, and solution concentration. This is particularly relevant in transportation applications, where decision-makers often require multiple feasible alternatives and operate under limited computational budgets.

\subsection{Parameter Selection and Experimental Settings}
The digitized annealing baseline was fixed to $10$ Trotter steps in order to balance physical resolution of the evolution with practical hardware limits. In preliminary transpilation, this choice produced circuits with two-qubit depth on the order of $10^3$, which is already substantial for current gate-based hardware given typical two-qubit error rates of approximately $6\times10^{-3}$. Using significantly more Trotter steps would therefore increase hardware cost beyond a practically meaningful regime for this study, while using substantially fewer steps would reduce the resolution of the annealing trajectory and weaken the value of the compression analysis.

The compression parameter $m$ was varied over $m\in\{0,\dots,6\}$. The upper limit $m=6$ was chosen because it corresponds to compressing a substantial fraction of the annealing circuit while still leaving a nontrivial tail for subsequent evolution or variational refinement. Since the problem Hamiltonian coefficients were normalized, this range spans approximately the first half of the effective annealing schedule, where compression is expected to be most reasonable. Compressing the entire circuit was not considered meaningful in this setting, both because later segments are more likely to contain gap-sensitive dynamics and because larger compressed targets require substantially greater classical preprocessing time in the AQC stage.

The QAOA depth was chosen as $p\in\{1,\dots,6\}$ to provide a direct comparison with the compression sweep and to remain within a regime that is appropriate for problems of this scale. For small QUBO instances such as those studied here, useful performance is typically obtained at relatively low QAOA depth, whereas larger $p$ mainly increases circuit depth and classical optimization burden.

\section{Results}

This section presents quantitative results for incremental AQC compression, hybrid QAOA replacement, linear-chain QAOA comparison, and spectral diagnostics. All experiments were performed on \verb|IBM_Rensselaer|, which has an Eagle processor with 127 qubits in a heavy-hex layout. The reported results should be interpreted as evaluating the ability of each circuit design to generate usable and diverse candidate plans under hardware constraints for each transportation problem. In essence, improvements in feasible sample counts correspond directly to higher probability of obtaining valid routing or assignment solutions from a single execution, while diversity reflects the range of alternatives available for downstream decision-making.

\subsection{Incremental Prefix Compression}

\begin{table*}[h]
\centering
\caption{Incremental AQC prefix compression for annealing circuits with 10 Trotter steps.}
\label{tab:aqc_prefix}
\begin{tabular}{cccccc}
\toprule
Problem & $m$ & \multicolumn{1}{l}{Prefix $D_{2q}$} & \multicolumn{1}{l}{Full $D_{2q}$} & \multicolumn{1}{l}{Feasible Samples} & \multicolumn{1}{l}{Unique Solutions} \\ \midrule
& 0   & 0               & 209                               & 64                                   & 30                                   \\
& 1   & 33             & 221                               & 79                                   & 32                                   \\
& 2   & 47                                  & 340                               & 76                                   & 31                                   \\
FLP & 3   & 39                                  & 182                               & 66                                   & 31                                   \\
& 4   & 69                                  & 186                               & 70                                   & 32                                   \\
&  5   & 96                                  & 282                               & 82                                   & 30                                   \\
& 6   & 78                                  & 144                               & 96                                   & 30                                   \\ 
\midrule
& 0 & 0   & 822  & 11 & 8 \\
& 1 & 108 & 1182 & 15 & 10 \\
& 2 & 129 & 837  & 20 & 14 \\
VRP & 3 & 182 & 941  &  8 & 7 \\
& 4 & 243 & 857  & 17 & 13 \\
& 5 & 266 & 877  & 16 & 13 \\
& 6 & 222 & 734  & 13 & 9 \\ 
\midrule

 & 0   & 0                                   & 1989                              & 3                                    & 3                                    \\
 & 1   & 153                                 & 1696                              & 4                                    & 4                                    \\
 & 2   & 213                                 & 1611                              & 2                                    & 2                                    \\
TSP  & 3   & 253                                 & 1556                              & 5                                    & 5                                    \\
 & 4   & 490                                 & 1598                              & 3                                    & 3                                    \\
 & 5   & 456                                 & 1271                              & 6                                    & 6                                    \\
 & 6   & 486                                 & 1368                              & 4                                    & 4                                    \\ 
\bottomrule
\end{tabular}
\end{table*}

Table~\ref{tab:aqc_prefix} reports the effect of progressively compressing larger prefixes of the annealing circuit using approximate quantum compilation (AQC), while leaving the remaining evolution unchanged. Results are shown across three representative transportation problems: FLP, VRP, and TSP. Across all cases, we observe a non-monotonic relationship between compression level $m$ and optimization performance.

Moderate prefix compression generally improves the probability of discovering feasible solutions while maintaining circuit depth comparable to or lower than the uncompressed case. For example, in the VRP instance, $m=2$ achieves the highest number of feasible samples (20) and the largest diversity (14 unique solutions). Similar trends are observed in FLP, where intermediate compression levels maintain strong feasibility performance, and in TSP, where modest compression improves the number of feasible tours sampled. These results suggest that early segments of the annealing trajectory contain redundant or smooth evolution that can be compressed without significantly altering the optimization dynamics.

However, more aggressive compression does not consistently improve performance. In VRP, increasing $m$ beyond moderate levels leads to a reduction in feasible samples despite comparable or even reduced circuit depth. In TSP, where feasibility is more constrained, performance is particularly sensitive to compression, with fluctuations in the number of feasible tours across $m$. This behavior is consistent with the structure of adiabatic evolution, where intermediate portions of the interpolation often contain avoided crossings or narrow spectral gaps. Compression that distorts these gap-sensitive regions can alter the state trajectory and degrade optimization performance.

Despite this, the highest compression level ($m=6$) consistently yields the smallest overall circuit depth across all problem classes, while still producing feasible solutions in every case. This indicates that compression primarily affects the distribution of probability mass over solutions rather than eliminating feasible solutions entirely.

Overall, these results support the hypothesis that early portions of the Trotterized annealing path are more compressible than later, gap-critical regions. Partial prefix compression therefore provides a practical mechanism for reducing circuit depth while preserving feasibility and access to high-quality solutions, highlighting a fundamental design tradeoff in near-term quantum optimization: deeper circuits improve solution concentration, while compressed circuits enhance hardware feasibility and maintain diverse candidate generation.

\subsection{Problem-Dependent Effects of Prefix Compression in Transportation Optimization}

\begin{figure}
    \centering

    \begin{subfigure}[b]{\linewidth}
        \centering
        \includegraphics[width=\linewidth]{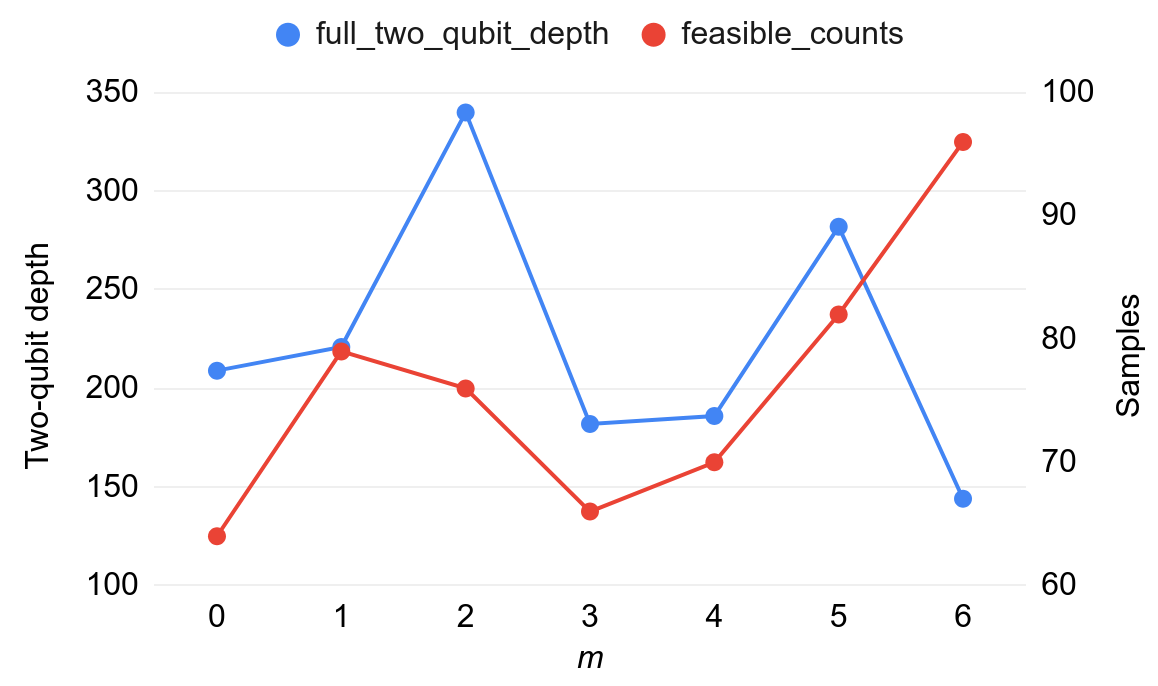}
        \caption{FLP}
        \label{fig:flp_anneal}
    \end{subfigure}

    \begin{subfigure}[b]{\linewidth}
        \centering
        \includegraphics[width=\linewidth]{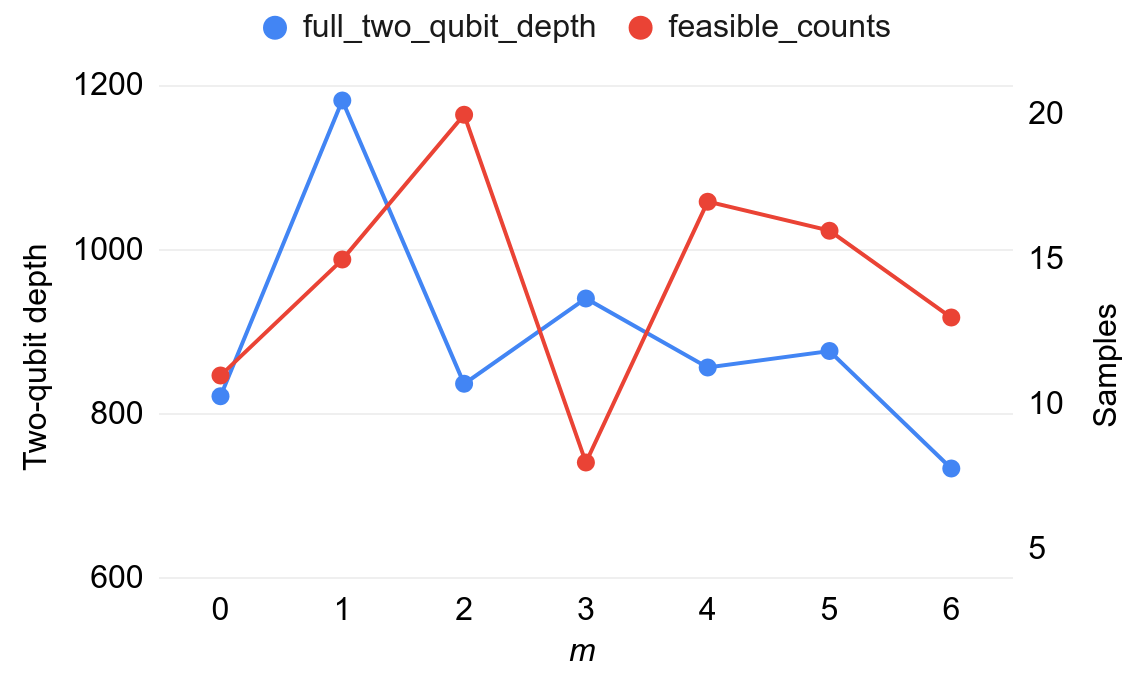}
        \caption{VRP}
        \label{fig:vrp_anneal}
    \end{subfigure}

    \begin{subfigure}[b]{\linewidth}
        \centering
        \includegraphics[width=\linewidth]{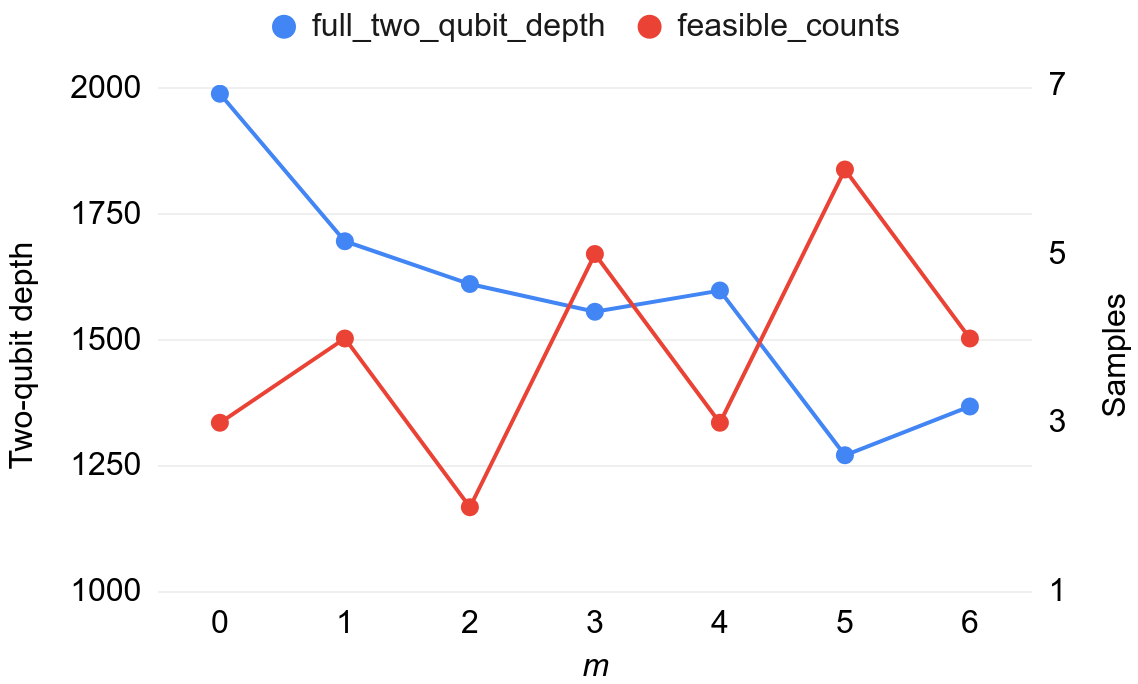}
        \caption{TSP}
        \label{fig:tsp_anneal}
    \end{subfigure}

    \caption{Two-qubit circuit depth (left axis) and number of feasible solutions (right axis) as a function of compressed prefix length $m$ for annealing circuits across transportation problem classes: (a) facility location, (b) vehicle routing, and (c) traveling salesman. The results illustrate the tradeoff between circuit depth and feasibility, showing that moderate prefix compression can reduce hardware cost while maintaining or improving feasible-solution discovery. }
    \label{fig:anneal_cross_problem}
\end{figure}

Figures~\ref{fig:flp_anneal}--\ref{fig:tsp_anneal} compare prefix-compression performance across three transportation optimization models: an uncapacitated facility location problem (FLP) with 2 facilities and 5 customers, yielding a 12-variable QUBO; a vehicle routing problem (VRP) with 4 nodes and 2 vehicles, yielding a 14-variable QUBO; and a traveling salesman problem (TSP) with 4 nodes, yielding a 16-variable QUBO. These instances span strategic siting and assignment, depot-based routing, and tour construction, respectively, and provide a compact testbed for examining how compressed annealing prefixes affect both compiled two-qubit depth and feasible-solution discovery across different transportation problem structures.

Across the three problem classes, the figure shows that prefix compression does not have a uniform effect on hardware cost or feasible-solution discovery. For FLP, feasible sampling generally improves as the compressed prefix length increases, with the largest feasible count attained at the most aggressive compression level shown. At the same time, the two-qubit depth varies non-monotonically, indicating that stronger prefix compression does not automatically translate into a shallower compiled circuit. VRP exhibits a similar pattern: feasible counts improve at selected intermediate values of $m$, especially around $m=2,4$, and 5, but the total two-qubit depth remains relatively high and irregular across the sweep. These results suggest that, for FLP and VRP, prefix compression can still reshape the prepared state in a way that benefits feasible solution discovery, even when it does not consistently reduce final compiled depth.

A likely reason is that for VRP and FLP the underlying problem Hamiltonians induce denser or less structured effective interactions after mapping and transpilation. As a result, the compressed prefix may still compile into circuits with substantial routing overhead, limiting the depth savings achievable through AQC alone. In contrast, the TSP instance shows clearer depth reduction as $m$ increases, while feasible counts remain within a comparable range and even improve at some intermediate and larger compression levels. This behavior is consistent with the more regular row-and-column assignment structure of the TSP formulation, which appears to compress more naturally under the annealing-prefix scheme. In effect, the permutation-style encoding of TSP provides a circuit structure that is more amenable to compression, whereas the connectivity patterns arising in VRP and FLP reduce the extent to which prefix compression alone can lower hardware cost. From a transportation applications perspective, this comparison is useful because it shows that the value of compression depends not only on problem size, but also on formulation structure: assignment-style models such as TSP may benefit more directly from compilation-based depth reduction, while routing and facility-location models may require additional structural simplifications or hybrid replacements beyond prefix compression alone.

\begin{figure}
    \centering
    \begin{subfigure}[b]{0.8\linewidth}
        \centering
        \includegraphics[width=\linewidth]{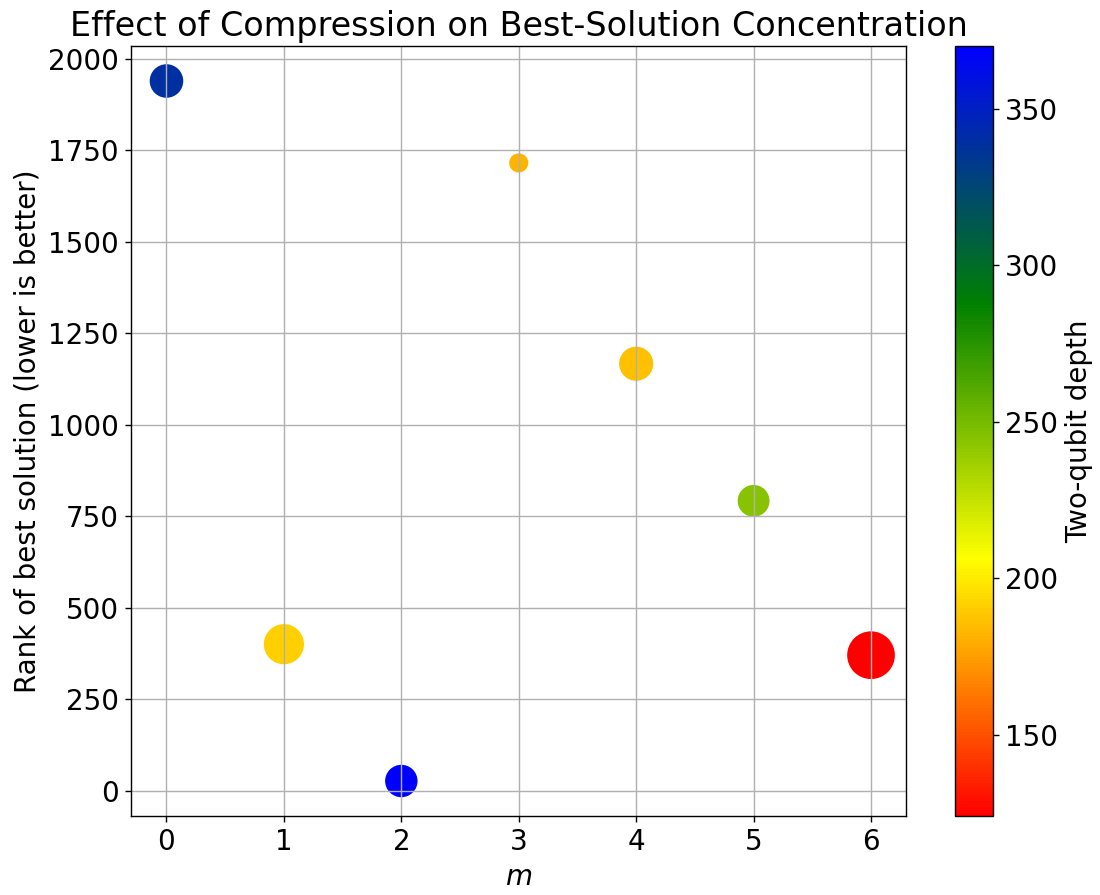}
        \caption{The ranking of the lowest cost assignment in the FLP at each compression level }
        \label{fig:rank_m_flp}
    \end{subfigure}

    \begin{subfigure}[b]{0.8\linewidth}
        \centering
        \includegraphics[width=\linewidth]{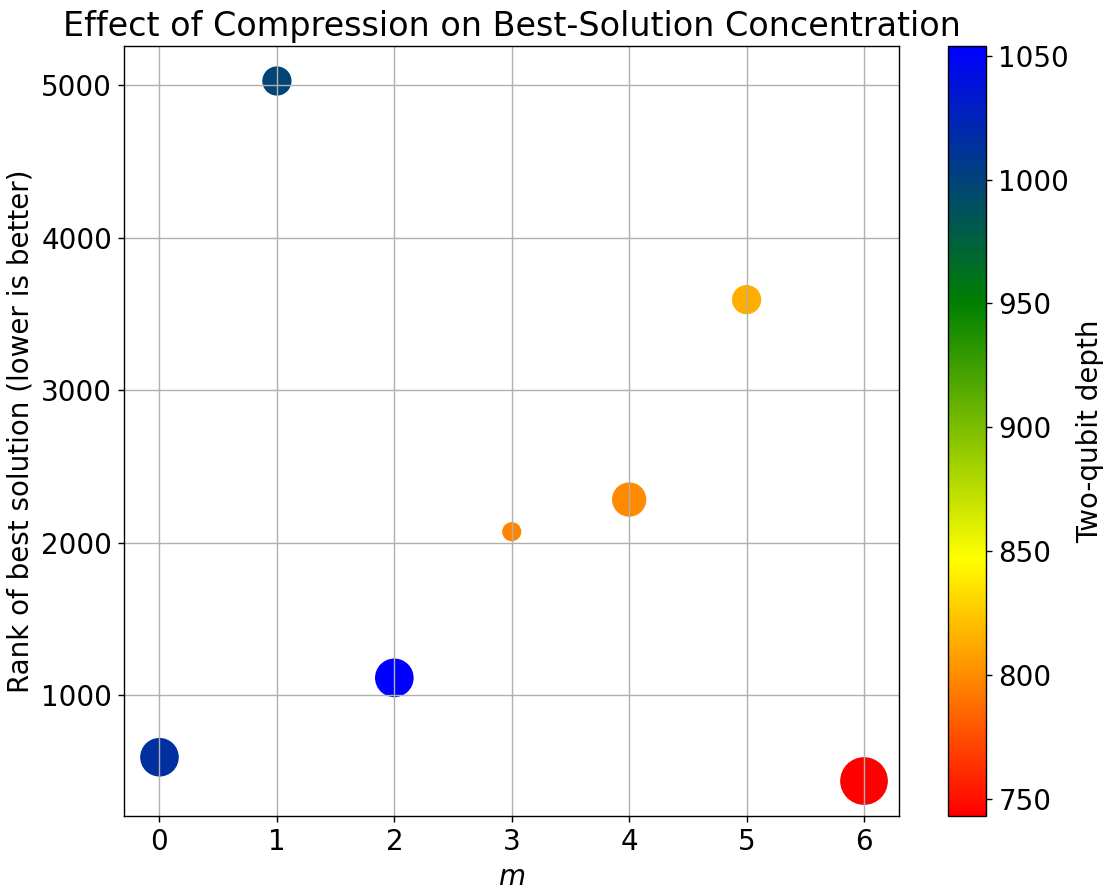}
        \caption{The ranking of the lowest cost routes for the VRP at each compression level}
        \label{fig:rank_m_vrp}
    \end{subfigure}

    \begin{subfigure}[b]{0.8\linewidth}
        \centering
        \includegraphics[width=\linewidth]{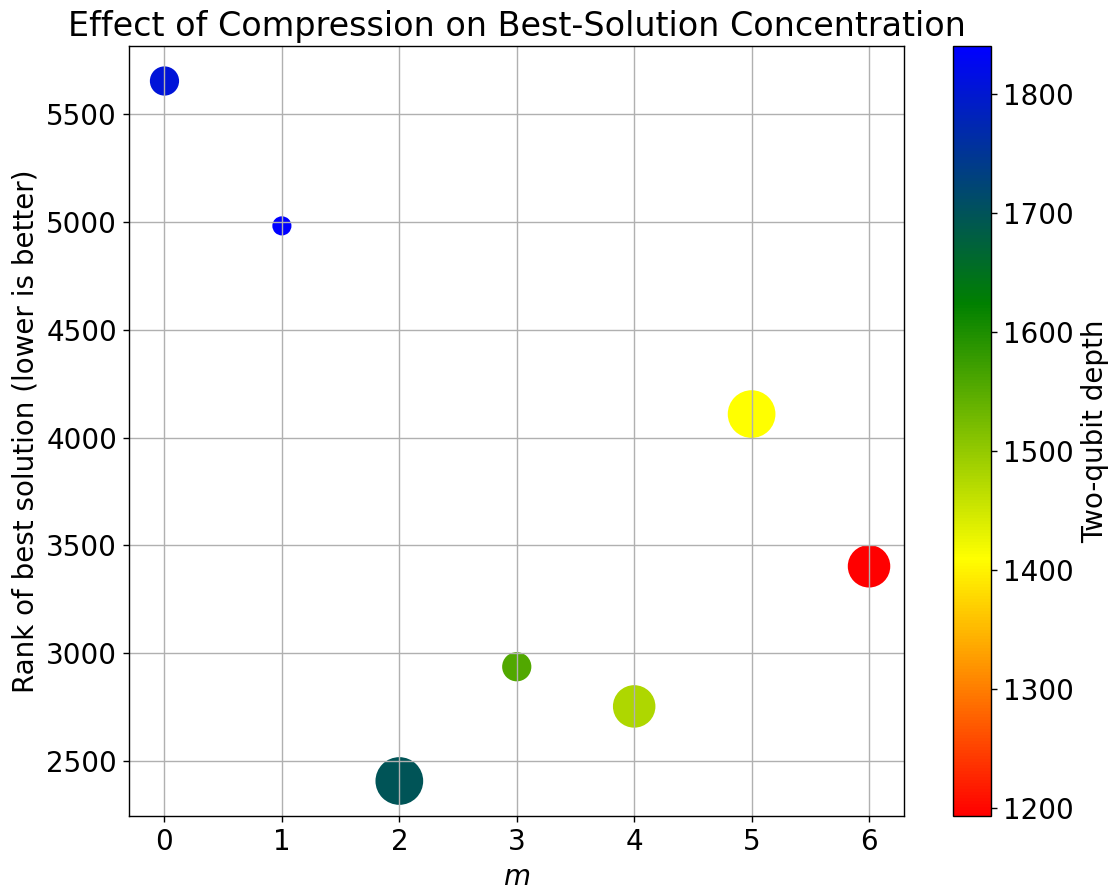}
        \caption{The ranking of the lowest cost tours for the TSP at each compression level}
        \label{fig:rank_m_tsp}
    \end{subfigure}

    \caption{Two-qubit depth is shown as color, and marker radius indicates the number of feasible solutions sampled from the circuit.}
    \label{fig:ranks_m}
\end{figure}

Figure \ref{fig:ranks_m} illustrates the effect of compression level $m$ on the sampling concentration of the best feasible solution across FLP, VRP, and TSP instances. The vertical axis reports the rank of the lowest-cost feasible solution among all sampled bitstrings (lower is better), marker size indicates the number of feasible solutions sampled, and color encodes the transpiled two-qubit gate depth.

Across all problem classes, a consistent trend emerges: increasing compression (larger $m$) significantly reduces circuit depth while maintaining strong feasibility performance. In particular, the highest compression level ($m=6$) yields the smallest two-qubit depth in all cases and continues to produce a substantial number of feasible solutions, as indicated by the relatively large marker sizes. This suggests that aggressive compression does not collapse the feasible subspace, but instead preserves the algorithm’s ability to generate usable candidate solutions.

However, maximal compression does not always yield the best sampling concentration of the optimal solution. Intermediate compression levels (e.g., $m=2$) often achieve lower ranks, indicating that the optimal solution appears more prominently among the most frequently sampled bitstrings. This reflects a tradeoff between circuit depth and probability concentration: deeper circuits may better concentrate probability mass on high-quality solutions, while more compressed circuits distribute probability more broadly across the feasible space.

Importantly, in all three problem instances, the true optimal solution is still sampled at the highest compression level. This indicates that compression primarily affects the distribution of probability mass rather than eliminating high-quality solutions from the support of the sampled distribution. From a transportation perspective, this is desirable: even if the optimal solution is not the most probable, it remains accessible within a finite sampling budget, while the reduced circuit depth improves hardware feasibility.

Overall, these results highlight a key design tradeoff in hybrid AQC–QAOA methods: higher compression improves hardware efficiency and maintains diverse feasible solution generation, while moderate compression can enhance the prominence of optimal solutions. This reinforces the view of quantum optimization as a candidate-generation framework, where algorithm design balances depth, feasibility, and solution concentration under NISQ constraints.

\subsection{Susceptibility}

\begin{figure}
    \centering
    \begin{subfigure}[t]{1.0\linewidth}
    \centering
    \includegraphics[width=\linewidth]{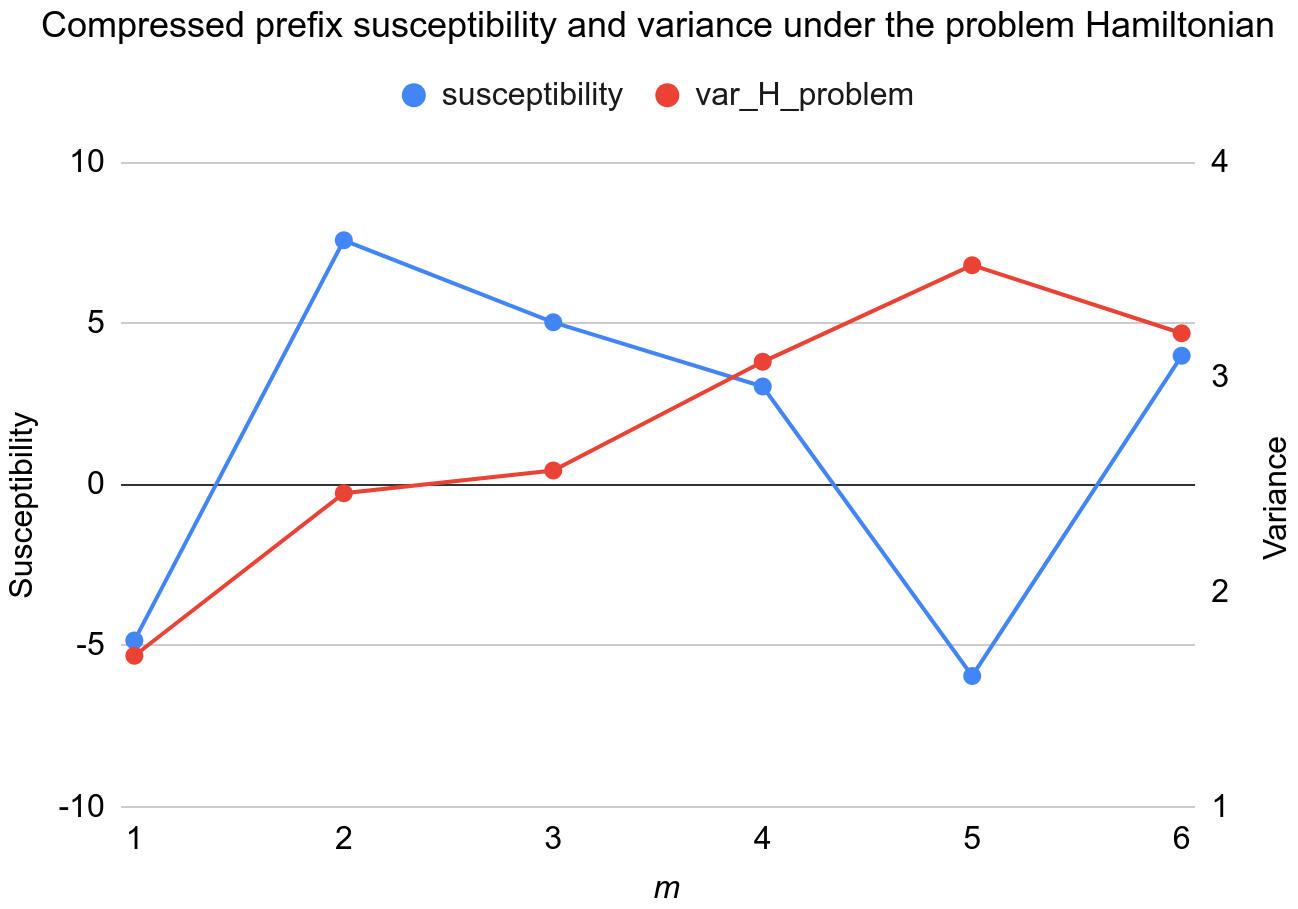}
    \caption{Cost variance under the problem Hamiltonian and perturbative susceptibility of the compressed annealing prefixes as a function of prefix length $m$. Peaks in these diagnostics indicate regions where the state becomes sensitive to perturbations and where the spectral gap is expected to be small.}
    \label{fig:prefix_spectral}
    \end{subfigure}

    \begin{subfigure}[t]{1.0\linewidth}
    \centering
    \includegraphics[width=\linewidth]{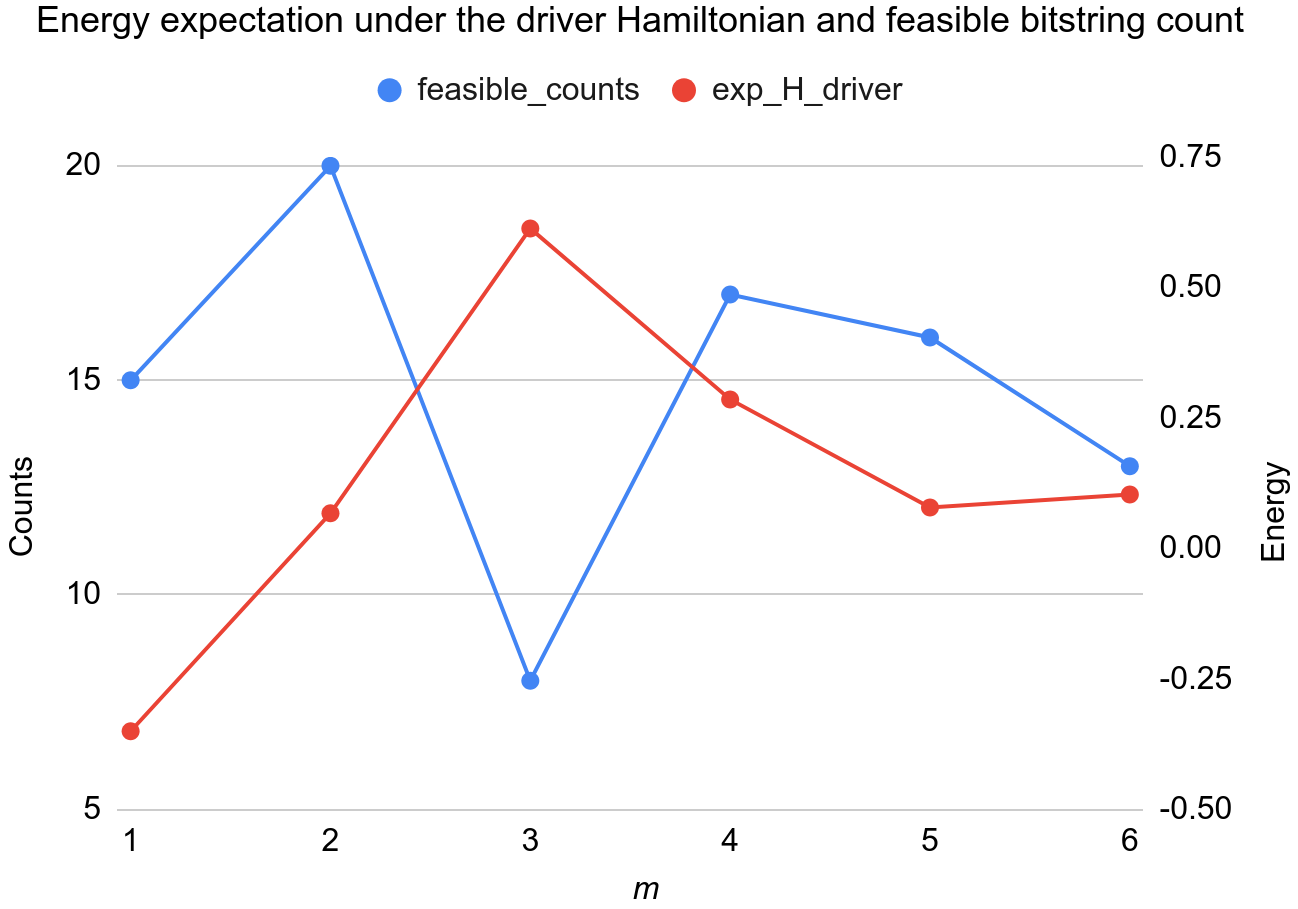}
    \caption{Relationship between the driver Hamiltonian expectation value of the compressed prefix state and the number of feasible solutions sampled by the full hybrid circuit. The driver energy $\langle H_B\rangle$ is measured after the prefix of length $m$, while feasible counts are obtained after executing the full circuit (prefix + tail). Lower driver energy corresponds to states that have rotated further away from the initial driver ground state and toward the cost Hamiltonian basis.}
    \label{fig:expDriver_feas}
    \end{subfigure}
        \caption{Diagnostic analysis of compressed annealing prefixes for the VRP instance. }
    \label{fig:prefix_diagnostics}
\end{figure}

Figure~\ref{fig:prefix_spectral} shows two complementary diagnostics of the compressed annealing prefixes, the cost variance under the problem Hamiltonian $H_C$ and the perturbative susceptibility with respect to the driver observable $V=\sum_i X_i$. These quantities provide experimentally accessible proxies for spectral structure during the annealing evolution.

The cost (or energy) variance measures the spread of the prepared quantum state in the eigenbasis of the cost Hamiltonian. When the variance is small, the state is concentrated near a single eigenstate of $H_C$, whereas larger variance indicates that the state has support across multiple nearby eigenstates. In adiabatic quantum evolution, such spreading typically occurs in regions where the spectral gap between the ground state and excited states becomes small. As shown in the figure, the variance increases steadily with prefix length and reaches its maximum around $m=5$, suggesting that the intermediate stages of the Trotterized annealing trajectory correspond to the most spectrally complex region of the evolution.

The susceptibility provides a complementary diagnostic. It measures the sensitivity of the prepared state to a small perturbation of the Hamiltonian and is computed using a finite-difference estimate of the observable response. Large magnitudes of susceptibility indicate that the quantum state changes rapidly under small perturbations, which commonly occurs near avoided crossings or narrow spectral gaps. The susceptibility curve exhibits pronounced extrema near $m=2$ and $m=5$, indicating that these prefixes correspond to regions where the spectral structure of the Hamiltonian changes most rapidly.

These diagnostics suggest that the compressed prefixes capture the gap-sensitive regions of the annealing trajectory where the system transitions between different dominant eigenstates. These regions are particularly important for optimization algorithms because they correspond to stages where the quantum state begins to acquire information about the cost landscape. The results therefore indicate that AQC compression preserves the physically important portions of the adiabatic path, allowing the hybrid algorithm to retain the spectral structure that guides the search toward low-cost solutions while substantially reducing circuit depth.

Figure~\ref{fig:expDriver_feas} shows the relationship between the expectation value of the driver Hamiltonian after the compressed prefix and the number of feasible solutions obtained from the full hybrid circuit. The quantity $\langle H_B \rangle$, where $H_B=\sum_i X_i$, measures how strongly the prefix state remains aligned with the initial transverse-field ground state $|+\rangle^{\otimes n}$. Larger values of $\langle H_B \rangle$ indicate that the state retains strong driver structure and remains close to a uniform superposition over bitstrings, while smaller values indicate that the state has rotated toward the eigenbasis of the problem Hamiltonian.

An approximate anticorrelation is observed between the driver energy of the prefix and the number of feasible solutions produced by the complete circuit, particularly with the most salient peak at $m=3$. Prefixes with larger $\langle H_B \rangle$ tend to yield fewer feasible solutions, whereas prefixes with reduced driver energy generally lead to improved feasible-solution discovery. This suggests that prefixes which partially break the driver symmetry and incorporate information from the problem Hamiltonian provide more effective initialization for the subsequent variational layers. In the hybrid circuit, the prefix therefore acts as a physics-informed preparation stage that biases the state toward regions of the solution space containing feasible transportation solutions, allowing the QAOA tail to refine these candidates rather than discovering them from an essentially uniform initial distribution.

\subsection{Hybrid Replacement with Standard QAOA}

\begin{table}[!htbp]
\centering
\caption{AQC compressed prefix + QAOA performance across compression depth $m$ and QAOA layers $p$ for the 4-node VRP.}
\label{tab:hybrid_qaoa}
\begin{tabular}{cccccc}
\hline
$m$ & $p$ & 2Q Depth & Iterations &  \multicolumn{1}{l}{\begin{tabular}[c]{@{}l@{}}Feasible \\ Routes\end{tabular}} &  \multicolumn{1}{l}{\begin{tabular}[c]{@{}l@{}}Unique \\ Solutions\end{tabular}} \\ \hline
0 & 1 & 105 & 53  & 23 & 12 \\
0 & 2 & 214 & 78  & 11 & 7  \\
0 & 3 & 349 & 108 & 21 & 13 \\
0 & 4 & 339 & 134 & 12 & 10 \\
0 & 5 & 538 & 160 & 12 & 9  \\
0 & 6 & 779 & 196 & 13 & 11 \\ \hline

1 & 1 & 208 & 46 & 11 & 8 \\
1 & 2 & 281 & 76 & 15 & 13 \\
1 & 3 & 422 & 106 & 14 & 9 \\
1 & 4 & 429 & 152 & 10 & 8 \\
1 & 5 & 701 & 160 & 10 & 8 \\
1 & 6 & 811 & 188 & 24 & 15 \\ \hline

2 & 1 & 234 & 50 & 18 & 14 \\
2 & 2 & 384 & 78 & 17 & 12 \\
2 & 3 & 435 & 105 & 19 & 15 \\
2 & 4 & 494 & 132 & 11 & 9 \\
2 & 5 & 562 & 160 & 11 & 9 \\
2 & 6 & 829 & 190 & 9 & 7 \\ \hline

3 & 1 & 255 & 48 & 18 & 10 \\
3 & 2 & 370 & 78 & 15 & 13 \\
3 & 3 & 383 & 106 & 15 & 12 \\
3 & 4 & 558 & 132 & 13 & 11 \\
3 & 5 & 668 & 166 & 10 & 9 \\
3 & 6 & 809 & 190 & 18 & 14\\ \hline

4 & 1 & 386 & 53  & 15 & 14 \\
4 & 2 & 401 & 77  & 14 & 11 \\
4 & 3 & 653 & 104 & 17 & 13 \\
4 & 4 & 684 & 132 & 13 & 10 \\
4 & 5 & 835 & 160 & 14 & 12 \\
4 & 6 & 824 & 188 & 12 & 9 \\ \hline

5 & 1 & 303 & 44  & 15 & 11 \\
5 & 2 & 482 & 76  & 13 & 10 \\
5 & 3 & 542 & 104 & 17 & 15 \\
5 & 4 & 653 & 133 & 12 & 10 \\
5 & 5 & 735 & 160 & 21 & 17 \\
5 & 6 & 976 & 190 & 9  & 8 \\ \hline

6 & 1 & 324 & 44  & 21 & 14 \\
6 & 2 & 501 & 78  & 19 & 13 \\
6 & 3 & 531 & 104 & 19 & 12 \\
6 & 4 & 645 & 134 & 14 & 10 \\
6 & 5 & 683 & 164 & 18 & 12 \\
6 & 6 & 920 & 188 & 16 & 14\\ \hline

\end{tabular}
\end{table}

\begin{figure}
    \centering
    \begin{subfigure}[t]{1.0\linewidth}
    \centering
    \includegraphics[width=\linewidth]{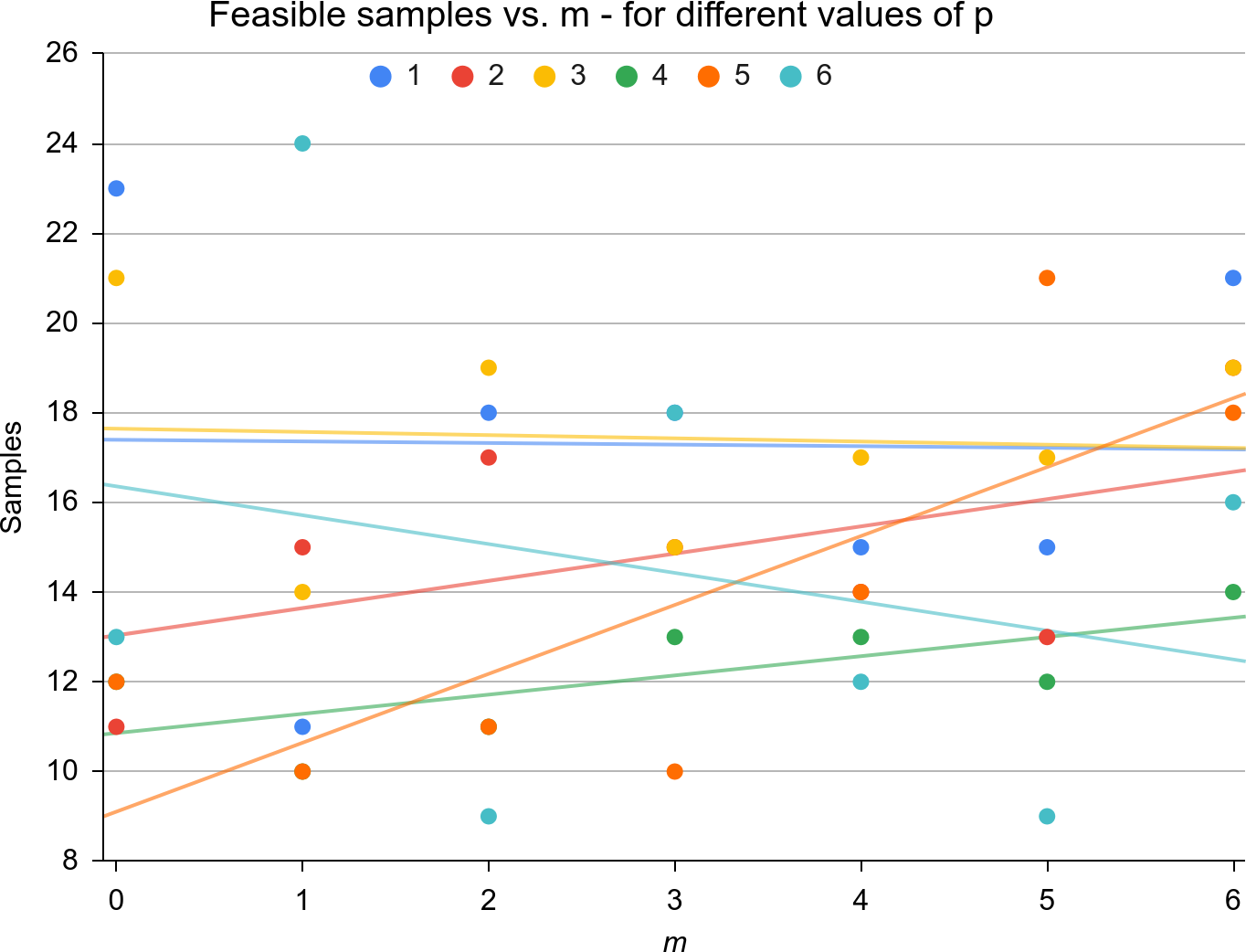}
    \caption{Total feasible routes as a function of AQC prefix length $m$ for different QAOA depths $p$. Points represents the number of feasible solutions obtained after optimization, while faded lines indicate trends for each $p$. Moderate compression generally improves feasible sample counts, though the relationship is non-monotonic and depends on the variational depth. }
    \label{fig:feas_vs_m_qaoa}
    \end{subfigure}

    \centering
    \begin{subfigure}[t]{1.0\linewidth}
    \centering
    \centering
    \includegraphics[width=\linewidth]{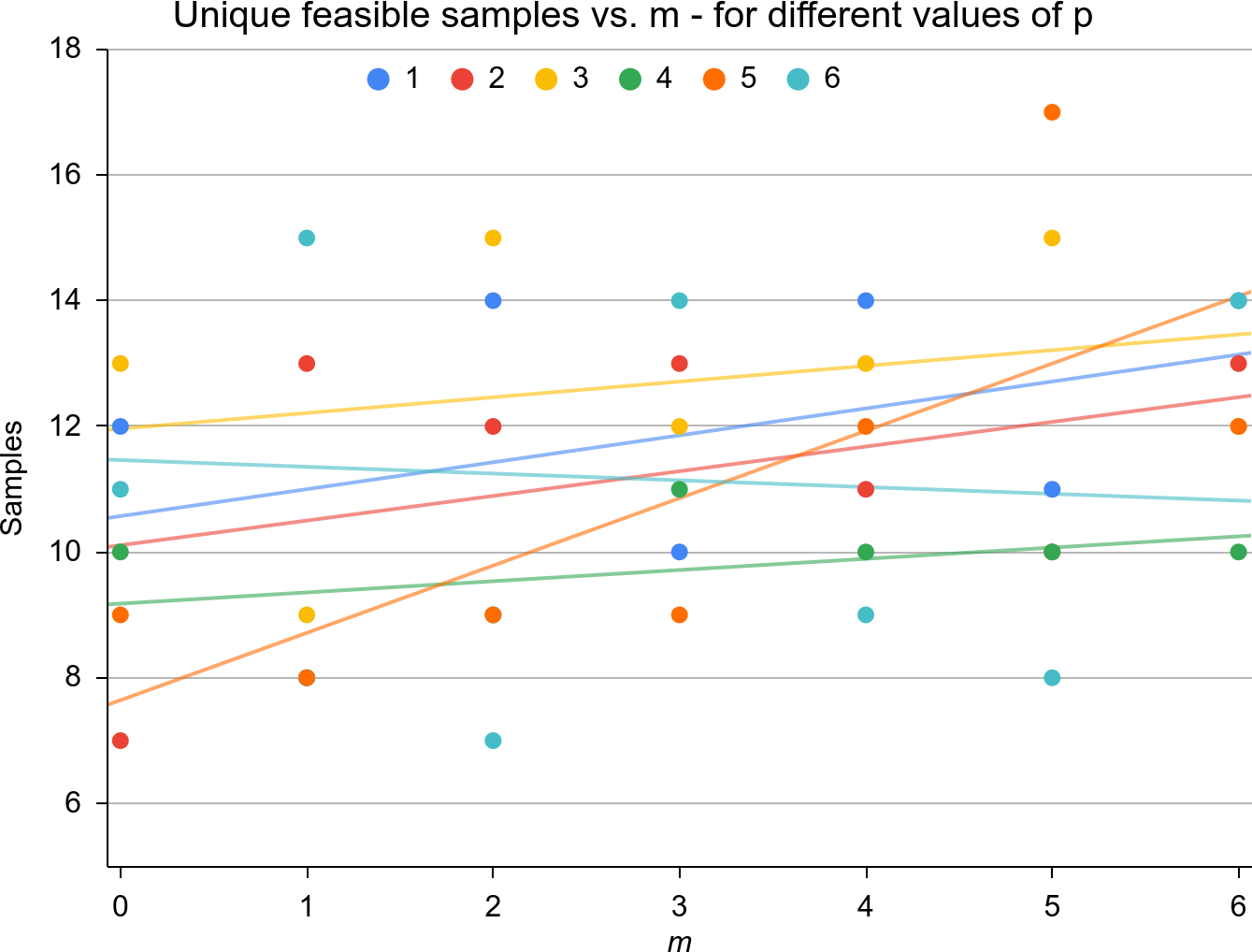}
    \caption{Number of unique feasible solutions as a function of AQC prefix length $m$ for different QAOA depths $p$. Each point denotes the number of distinct feasible routes observed, with faded lines showing trends for each $p$. Increasing compression can improve solution diversity for certain depths.}
    \label{fig:unique_vs_m_qaoa}
    \end{subfigure}
    
    \caption{Effect of AQC prefix compression on feasible solution generation from QAOA for the VRP instance. Moderate compression can improve feasibility and diversity, while aggressive compression shows mixed behavior across depths.}
    
\end{figure}

Figures~\ref{fig:feas_vs_m_qaoa} and~\ref{fig:unique_vs_m_qaoa} show the number of feasible routes and the diversity of those solutions discovered as a function of the AQC compressed prefix length $m$ for different QAOA depths $p$. The full information including two-qubit depth and number of iterations is shown in Table~\ref{tab:hybrid_qaoa}. Classical optimization of the parameters was carried out using the COBYQA algorithm via the minimize routine in scikit-learn.

Several trends are apparent. First, moderate levels of prefix compression generally maintain or improve the number of feasible solutions discovered. For many values of $p$, feasible routes increase or remain stable as $m$ grows from 0 to intermediate values. This suggests that portions of the early annealing trajectory can be compressed without significantly degrading the optimization landscape explored by the QAOA layers.

Second, deeper QAOA circuits appear to benefit more strongly from prefix compression. In particular, larger values of $p$ show an upward trend in feasible sampling as $m$ increases, indicating that replacing a portion of the annealing prefix with a compressed circuit may provide a better initialization for deeper variational optimization. This effect is especially visible for $p=4$ and $p=5$, where feasible route counts increase noticeably at larger values of $m$. In contrast, the trend reverses for $p=6$, which achieves its best performance at $m=1$. A possible explanation is the increased dimensionality of the variational optimization problem. At $p=6$, the circuit contains 12 free parameters, approaching the dimensionality of the underlying problem with 14 binary variables. In this regime the classical optimizer must navigate a larger parameter space, which may reduce the practical benefit of deeper QAOA layers.

Third, the diversity of feasible solutions closely tracks the overall feasible solution count. Regions of the parameter grid that produce more valid routes also tend to produce a larger number of unique feasible bitstrings. This indicates that the circuits are exploring a broader portion of the feasible solution space. In transportation optimization contexts, such diversity is valuable because it provides multiple candidate routing or facility configurations that may be useful for downstream decision analysis.

Overall, these results suggest that combining AQC prefix compression with QAOA can improve the exploration of feasible transportation solutions while avoiding the substantial circuit depth increases that would result from simply increasing the QAOA depth alone.

\begin{figure*}[t]
    \centering
    \begin{subfigure}[t]{0.49\linewidth}
    \centering
    \includegraphics[width=1.0\linewidth]{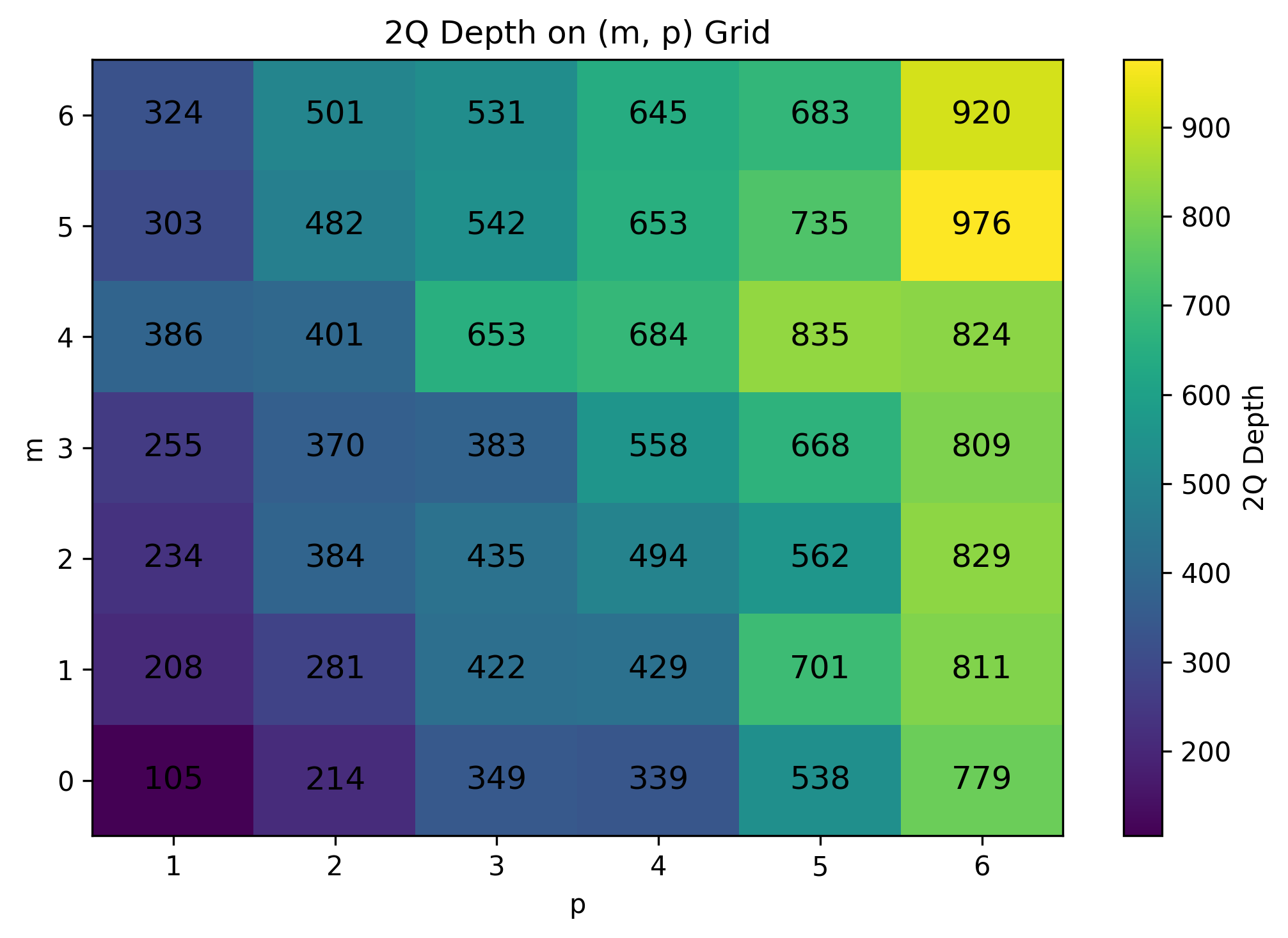}
    \caption{Transpiled two-qubit gate depth across the $(m,p)$ parameter grid. }
    \label{fig:heatmap_2q_qaoa}
    \end{subfigure}
\hfill
    \begin{subfigure}[t]{0.49\linewidth}
    \centering
    \includegraphics[width=1.0\linewidth]{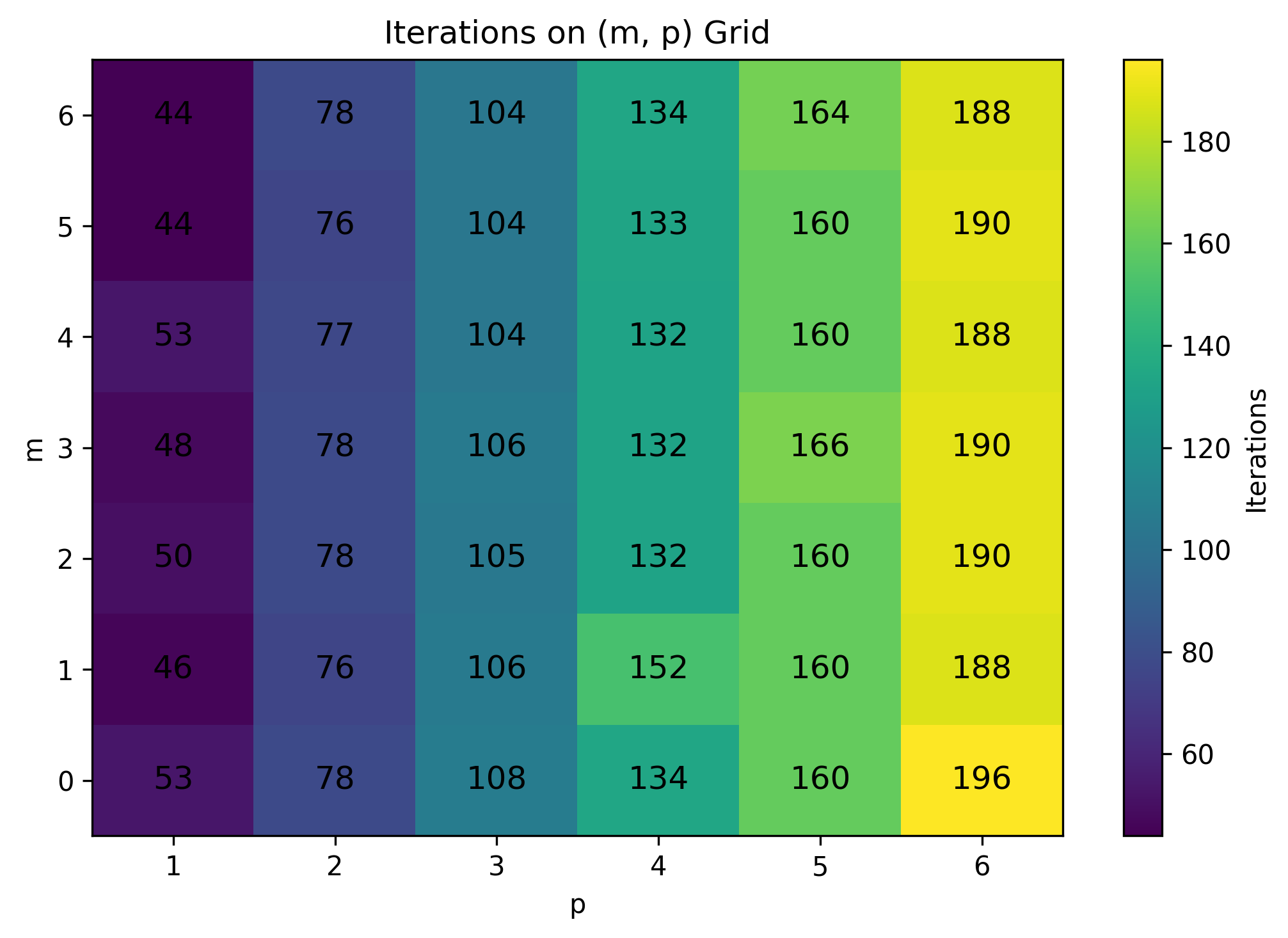}
    \caption{Classical optimization iterations required for convergence of the variational parameters.}
    \label{fig:heatmap_iter_qaoa}
    \end{subfigure}

    \begin{subfigure}[t]{0.49\linewidth}
    \centering
    \includegraphics[width=01.0\linewidth]{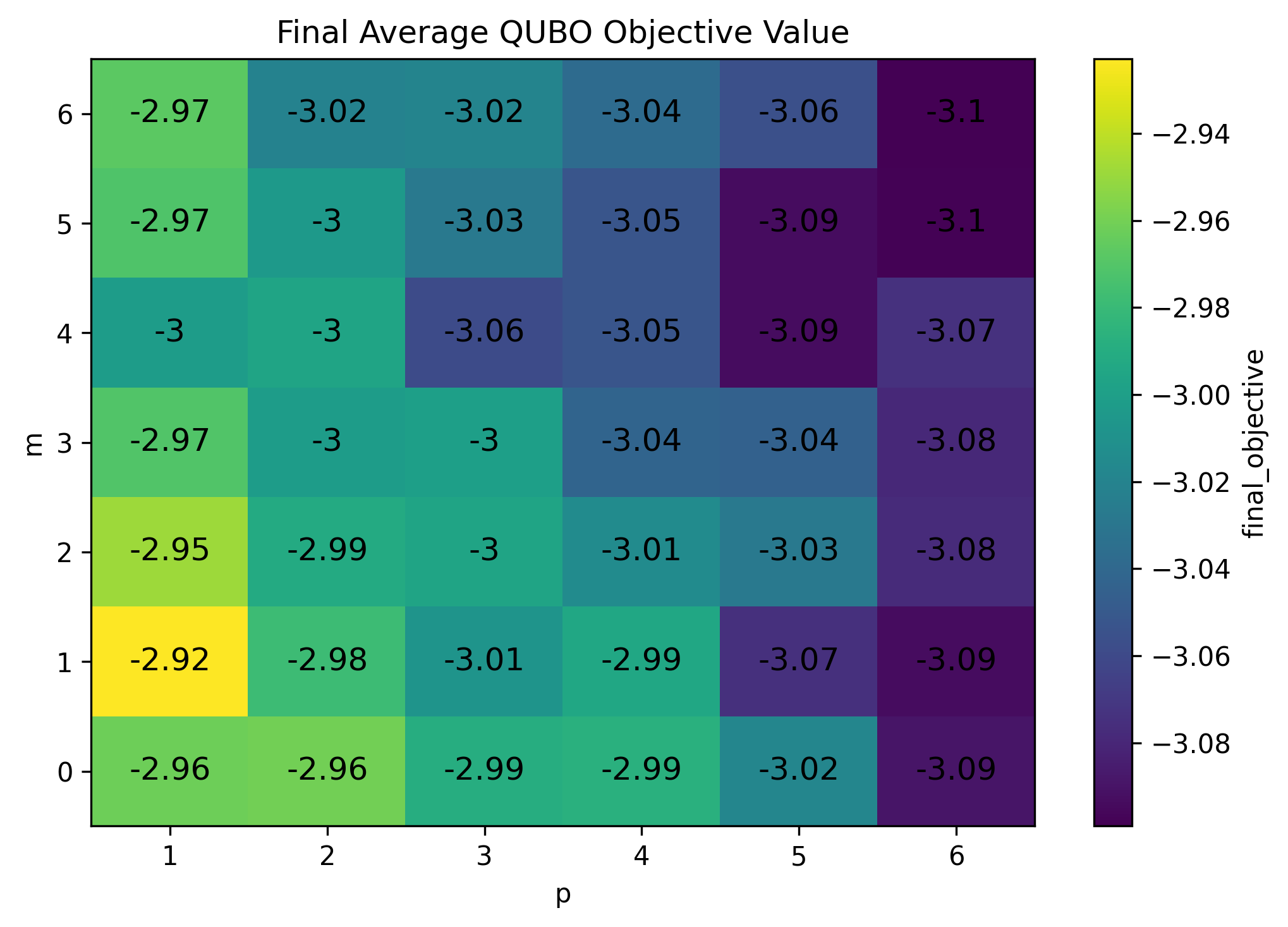}
    \caption{Final average objective value across the $(m,p)$ parameter grid. }
    \label{fig:obj_qaoa}
    \end{subfigure}
\hfill
    \begin{subfigure}[t]{0.49\linewidth}
    \centering
    \includegraphics[width=1.0\linewidth]{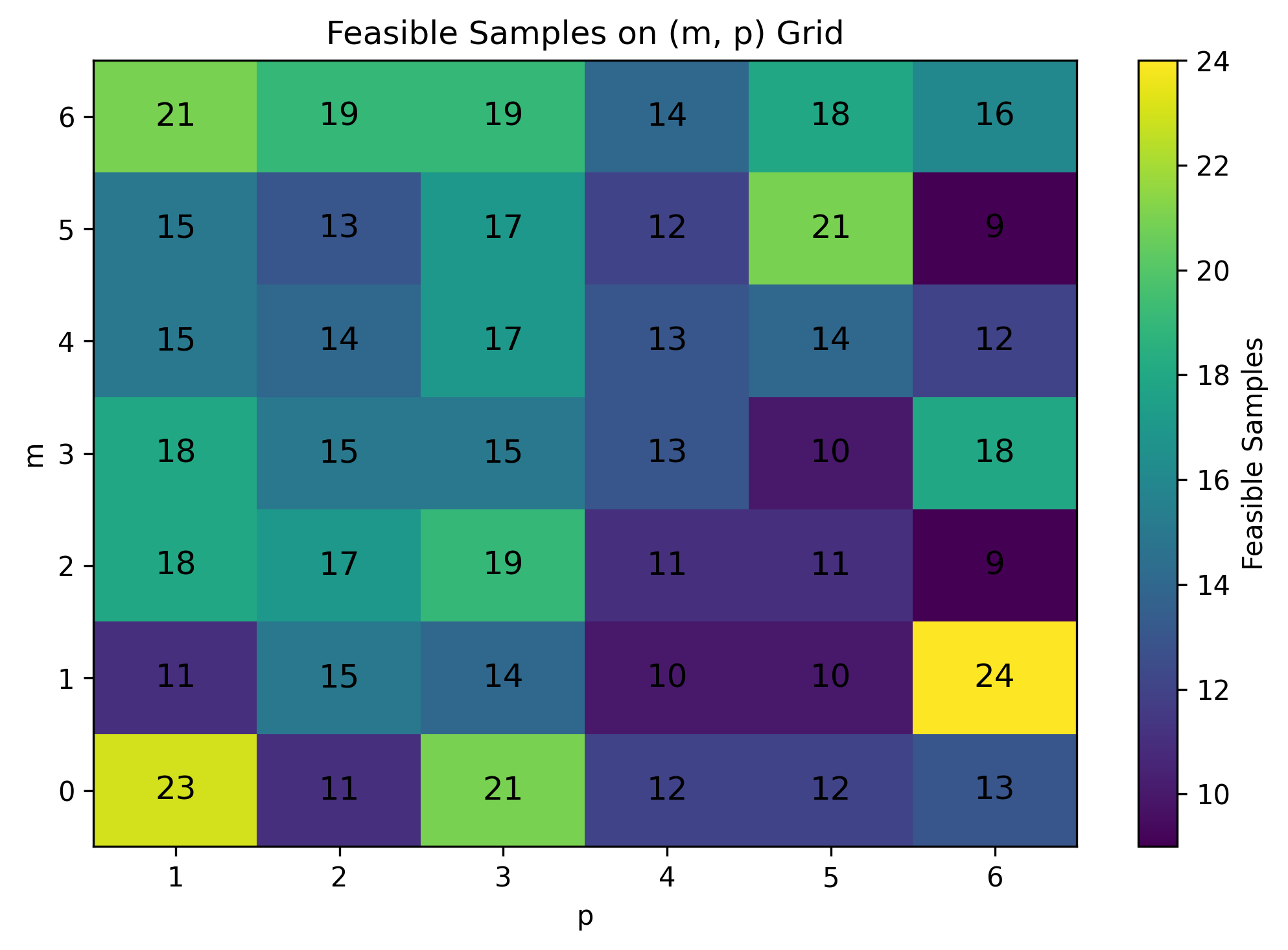}
    \caption{Feasible samples across the $(m,p)$ parameter grid.}
    \label{fig:heatmap_feasible_qaoa}
    \end{subfigure}
    \caption{Performance of the hybrid AQC+QAOA approach across the $(m,p)$ parameter grid. Results show tradeoffs between circuit depth, optimization effort, feasibility, and solution quality under varying prefix compression and QAOA depth.}
    \label{fig:heatmaps_qaoa}
\end{figure*}

Figure~\ref{fig:heatmaps_qaoa} summarizes the performance of the hybrid AQC+QAOA approach across the $(m,p)$ parameter grid for the VRP instance, highlighting the interplay between circuit depth, optimization effort, feasibility, and solution quality. 

The two-qubit depth heatmap in Figure~\ref{fig:heatmap_2q_qaoa}  shows that circuit cost is primarily driven by the QAOA depth $p$, increasing significantly as $p$ grows due to additional phase-separation and mixing layers. In contrast, variation with respect to the compressed prefix length $m$ is relatively modest, indicating that overall hardware cost is dominated by the variational QAOA component.
The optimization iterations heatmap in Figure~\ref{fig:heatmap_iter_qaoa} exhibits an even stronger dependence on $p$, with the number of classical iterations increasing steadily as $p$ grows, reflecting a more complex variational landscape. In contrast, iteration counts remain largely insensitive to $m$, suggesting that prefix compression mainly affects initialization rather than the intrinsic difficulty of the optimization.

A clear trend emerges from the depth landscape: configurations with full two-qubit depth below approximately 750 consistently achieve competitive objective values in Figure~\ref{fig:obj_qaoa}, despite deeper circuits showing the least expected values. In particular, moderate prefix compression combined with intermediate QAOA depth yields solutions close to the best observed costs, indicating that aggressive circuit depth is not required to reach high-quality routing plans.

More importantly from a transportation perspective, low-depth circuits also generate a large number of feasible solutions. As shown in Figure~\ref{fig:heatmap_feasible_qaoa}, configurations with reduced depth, especially those with moderate compression, maintain or even increase the number of feasible samples relative to deeper circuits. This is a critical property for vehicle routing applications, where decision-makers often require multiple valid route alternatives rather than a single optimized solution.

The results also reveal a tradeoff between optimization effort and circuit complexity. Higher QAOA depths increase the number of classical iterations required for convergence, without consistently improving solution quality, particularly in terms of feasible solutions. In contrast, shallower configurations reduce both circuit depth and optimization overhead while preserving feasibility and near-optimal costs.

Overall, these findings suggest that carefully compressed circuits operate in a regime that is both hardware-efficient and practically useful: they retain access to low-cost routing solutions while producing a diverse set of feasible candidates. This reinforces the view of quantum optimization as a candidate-generation tool for transportation systems, where feasibility, diversity, and computational efficiency are as important as optimality.

\subsection{Hybrid Replacement with Linear-Chain QAOA}

\begin{figure}
    \centering
    \begin{subfigure}[t]{1.0\linewidth}
    \centering
    \includegraphics[width=\linewidth]{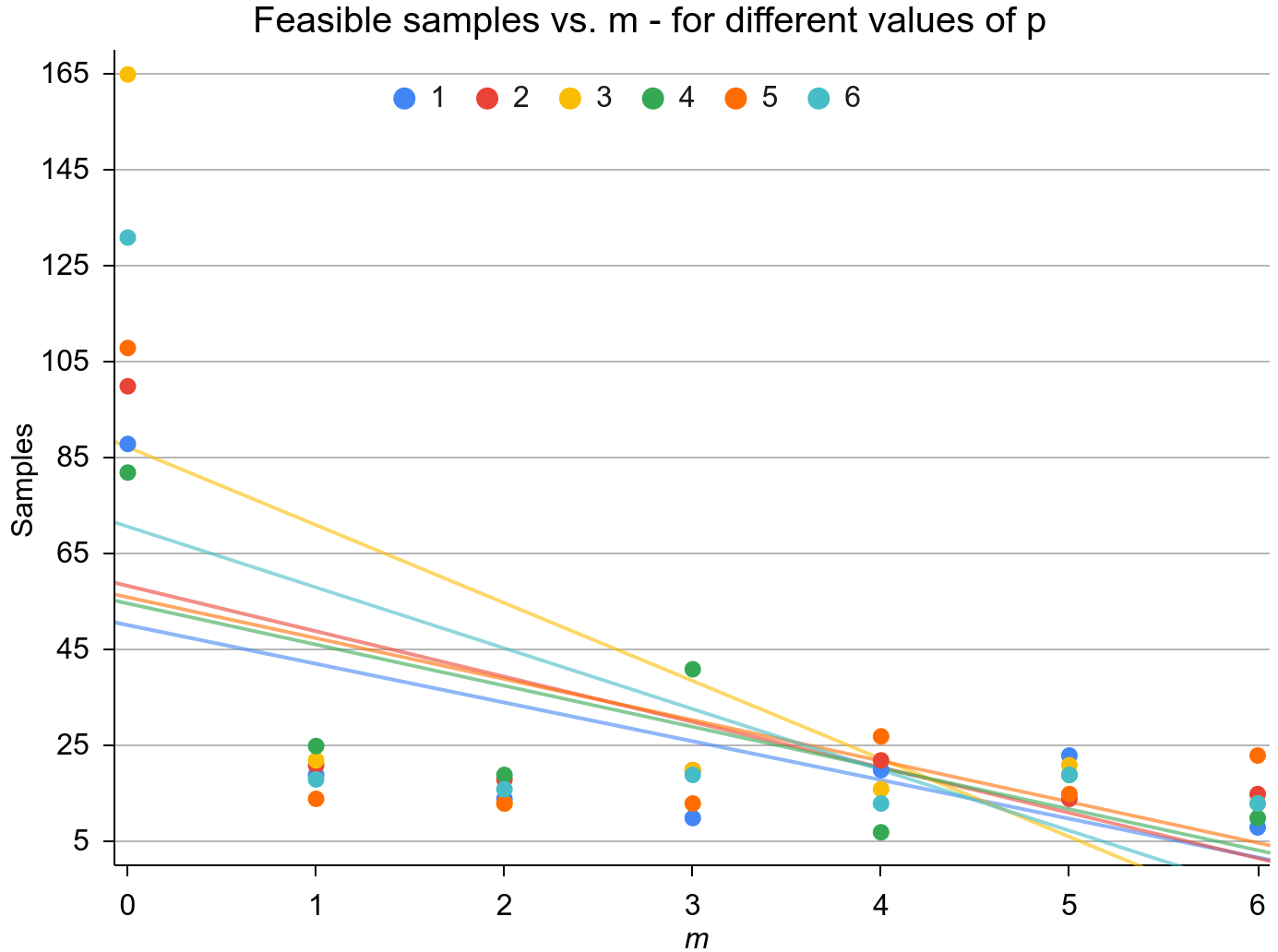}
    \caption{Feasible routes vs. AQC compressed prefix length $m$ across different values of $p$ layers of LC-QAOA. }
    \label{fig:feas_vs_m_lcqaoa}
    \end{subfigure}
\vspace{0.25cm}

    \begin{subfigure}[t]{1.0\linewidth}
    \centering
    \includegraphics[width=\linewidth]{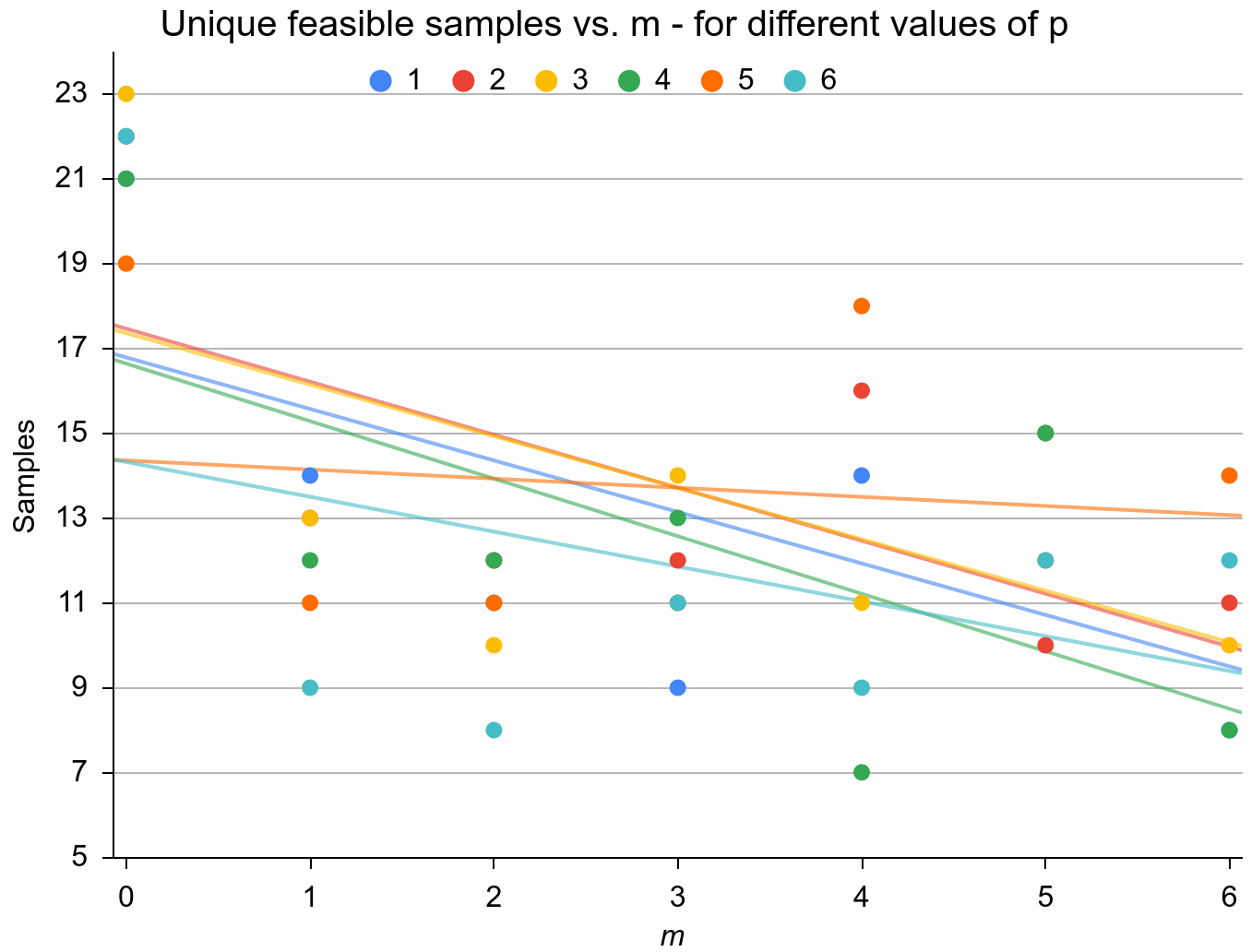}
    \caption{Unique feasible routes sampled vs. AQC compressed prefix length $m$ across different values of $p$ layers of LC-QAOA. }
    \label{fig:unique_vs_m_lcqaoa}
    \end{subfigure}
        \caption{AQC prefix compression on feasible solution generation with LC-QAOA for the VRP instance. The mismatch of the ans\"atz from adiabatic evolution causes a sharp drop in performance with compression. }
\end{figure}

Figures~\ref{fig:feas_vs_m_lcqaoa} and~\ref{fig:unique_vs_m_lcqaoa} show the number of feasible routes and the diversity of feasible solutions as functions of the AQC compressed prefix length $m$ for different LC-QAOA depths $p$. In this setting, LC-QAOA uses a locally truncated cost Hamiltonian retaining only adjacent $ZZ$ interactions. Full metrics, including circuit depth and optimization iterations, are reported in Table~\ref{tab:hybrid_lcqaoa}. The implementation is based on previous work \citep{azfar2026shallow}.

Unlike the standard QAOA results, feasible sampling generally decreases as $m$ increases. For most values of $p$, the highest feasible counts occur at $m=0$, after which performance declines with compressed prefix initialization. This indicates that LC-QAOA is more sensitive to perturbations introduced in the early annealing trajectory by prefix compression.

The dependence on $p$ also differs from the full-cost QAOA case. While larger $p$ improves feasible counts at low compression, these gains diminish rapidly as $m$ increases. This behavior can be attributed to the mismatch between the annealing prefix, generated using the full cost Hamiltonian, and the truncated Hamiltonian used in LC-QAOA. As a result, the variational layers no longer act as a continuation of the same dynamics and are less effective at refining the compressed state. Consequently, the LC-QAOA layers are not able to leverage the structure encoded in the compressed prefix,

This observation also highlights an important assumption underlying the AQC prefix compression approach: the compressed circuit is intended to approximate the early portion of the dynamics generated by the same Hamiltonian that appears in the subsequent variational layers. When the variational circuit instead has a different structure, as in the LC-QAOA formulation with a truncated cost operator, this continuity is broken and the compressed prefix may no longer provide a beneficial initialization for the optimization.

The diversity of feasible solutions closely mirrors the overall valid route counts. Configurations that produce more feasible solutions also yield a larger number of unique feasible bitstrings, indicating broader exploration of the feasible region. As $m$ increases, both feasible counts and solution diversity decline, suggesting that prefix reduces the algorithm's ability to explore alternative feasible transportation solutions.

Overall, these results show that AQC-based prefix initialization is not universally beneficial across variational algorithms. In the present setting, LC-QAOA performs best with no compressed prefix added, indicating that the advantages of annealing-based initialization do not automatically transfer to variational circuits built from truncated cost Hamiltonians. For transportation optimization, this highlights the importance of matching the initialization strategy to the structure of the downstream variational ans\"atz rather than assuming that prefix compression will improve every hybrid architecture.

\begin{table}[!htbp]
\centering
\caption{AQC compressed prefix + LC-QAOA performance across compression depth $m$ and QAOA layers $p$.}
\label{tab:hybrid_lcqaoa}
\begin{tabular}{cccccc}
\hline
$m$ & $p$ & 2Q Depth & Iterations &  \multicolumn{1}{l}{\begin{tabular}[c]{@{}l@{}}Feasible \\ Routes\end{tabular}} &  \multicolumn{1}{l}{\begin{tabular}[c]{@{}l@{}}Unique \\ Solutions\end{tabular}} \\ \hline
0   & 1  & 8  & 48   & 88 & 21  \\
0  & 2  & 12 & 87   & 100 & 22  \\
0  & 3  & 16 & 128  & 165 & 23  \\
0  & 4  & 20 & 134  & 82  & 21  \\
0  & 5  & 24 & 192  & 108 & 19  \\
0  & 6  & 28 & 214  & 131 & 22  \\ \hline
1  & 1  & 114     & 46   & 19 & 14  \\
1  & 2  & 135     & 77   & 21 & 13  \\
1  & 3  & 173     & 109  & 22 & 13  \\
1  & 4  & 166     & 139  & 25 & 12  \\
1  & 5  & 205     & 173  & 14 & 11  \\
1  & 6  & 195     & 189  & 18 & 9   \\ \hline
2  & 1  & 130     & 48   & 14 & 11  \\
2  & 2  & 189     & 87   & 18 & 12  \\
2  & 3  & 169     & 116  & 13 & 10  \\
2  & 4  & 183     & 134  & 19 & 12  \\
2  & 5  & 190     & 160  & 13 & 11  \\
2  & 6  & 184     & 194  & 16 & 8   \\ \hline
3  & 1  & 131     & 50   & 10 & 9   \\
3  & 2  & 131     & 78   & 20 & 12  \\
3  & 3  & 187     & 109  & 20 & 14  \\
3  & 4  & 133     & 135  & 41 & 13  \\
3  & 5  & 183     & 163  & 13 & 11  \\
3  & 6  & 122     & 192  & 19 & 11  \\ \hline
4  & 1  & 279     & 51   & 20 & 14  \\
4  & 2  & 279     & 78   & 22 & 16  \\
4  & 3  & 279     & 111  & 16 & 11  \\
4  & 4  & 279     & 133  & 7  & 7   \\
4  & 5  & 279     & 183  & 27 & 18  \\
4  & 6  & 279     & 192  & 13 & 9   \\ \hline
5  & 1  & 257     & 51   & 23 & 15  \\
5  & 2  & 257     & 78   & 14 & 10  \\
5  & 3  & 257     & 105  & 21 & 15  \\
5  & 4  & 257     & 135  & 19 & 15  \\
5  & 5  & 257     & 159  & 15 & 12  \\
5  & 6  & 257     & 188  & 19 & 12  \\ \hline
6  & 1  & 273     & 47   & 8  & 8   \\
6  & 2  & 273     & 82   & 15 & 11  \\
6  & 3  & 273     & 103  & 13 & 10  \\
6  & 4  & 273     & 133  & 10 & 8   \\
6  & 5  & 273     & 167  & 23 & 14  \\
6  & 6  & 273     & 192  & 13 & 12 \\ \hline

\end{tabular}
\end{table}

\begin{figure*}[t]
    \centering
    \begin{subfigure}[t]{0.49\linewidth}
    \centering
    \includegraphics[width=1.0\linewidth]{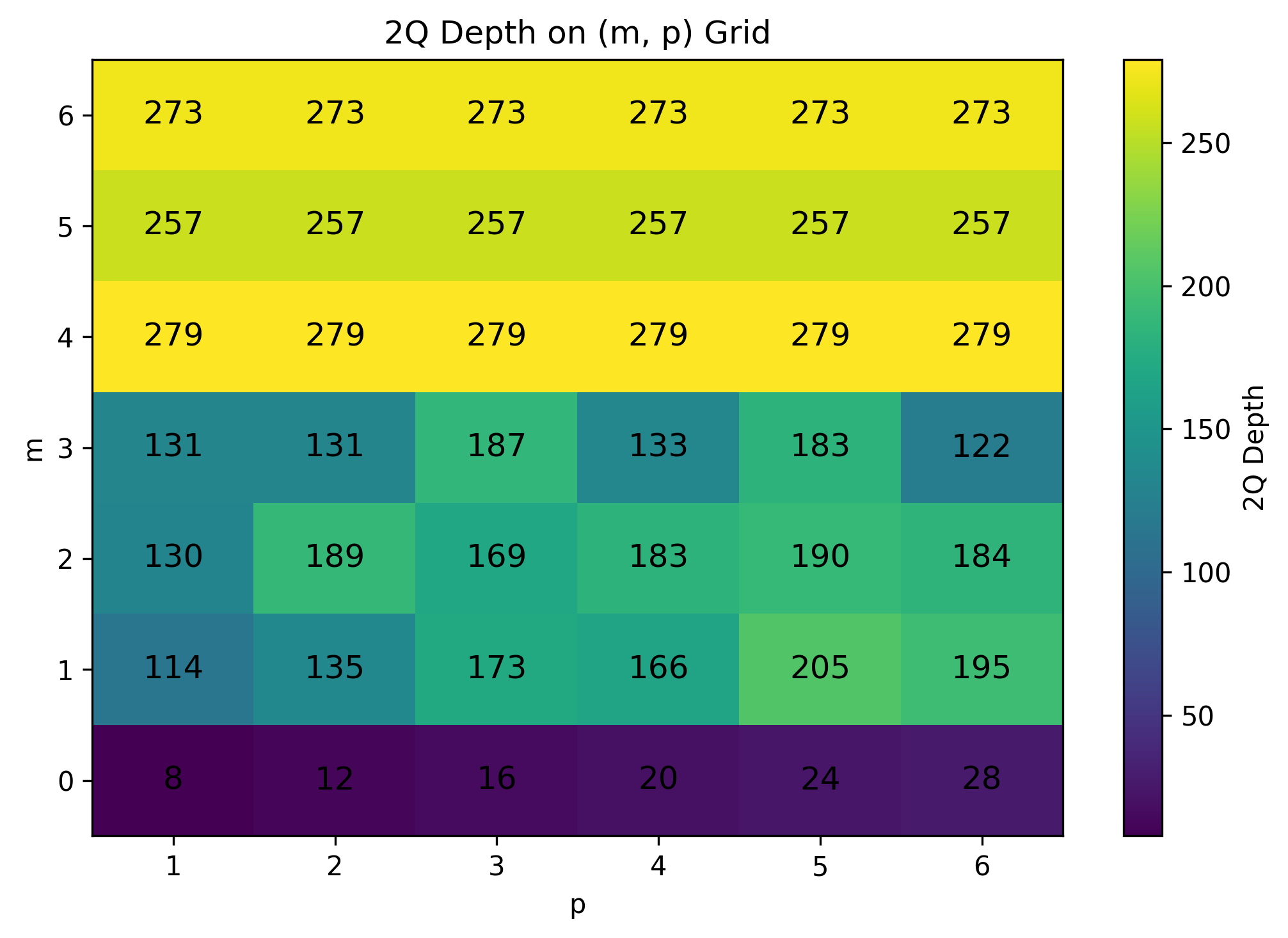}
    \caption{Transpiled two-qubit gate depth across the $(m,p)$ parameter grid.}
    \label{fig:heatmap_2q_lcqaoa}
    \end{subfigure}
\hfill
    \begin{subfigure}[t]{0.49\linewidth}
    \centering
    \includegraphics[width=1.0\linewidth]{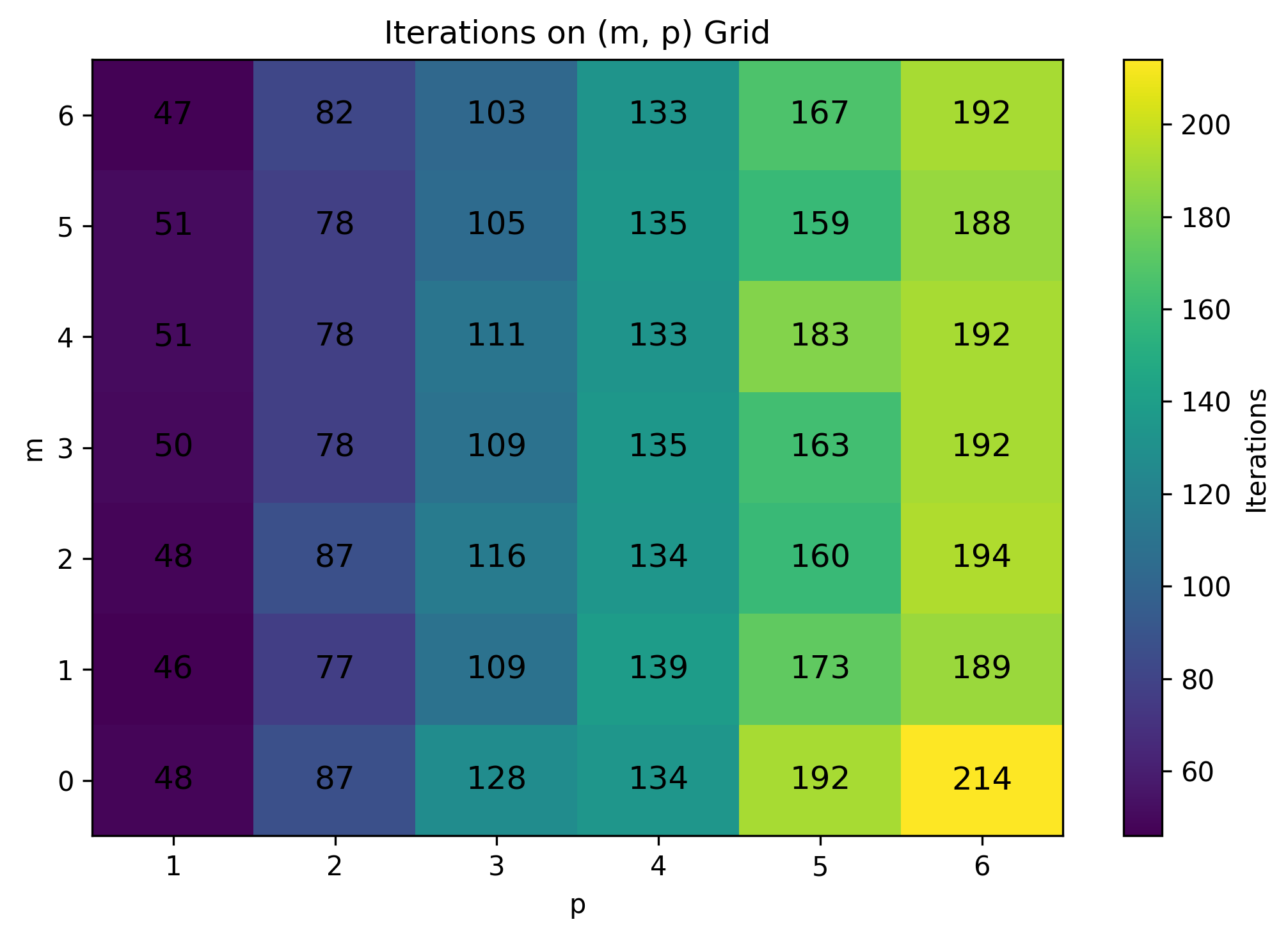}
    \caption{Classical optimization iterations required for convergence of the variational parameters.}
    \label{fig:heatmap_iter_lcqaoa}
    \end{subfigure}

    \begin{subfigure}[t]{0.49\linewidth}
    \centering
    \includegraphics[width=1.0\linewidth]{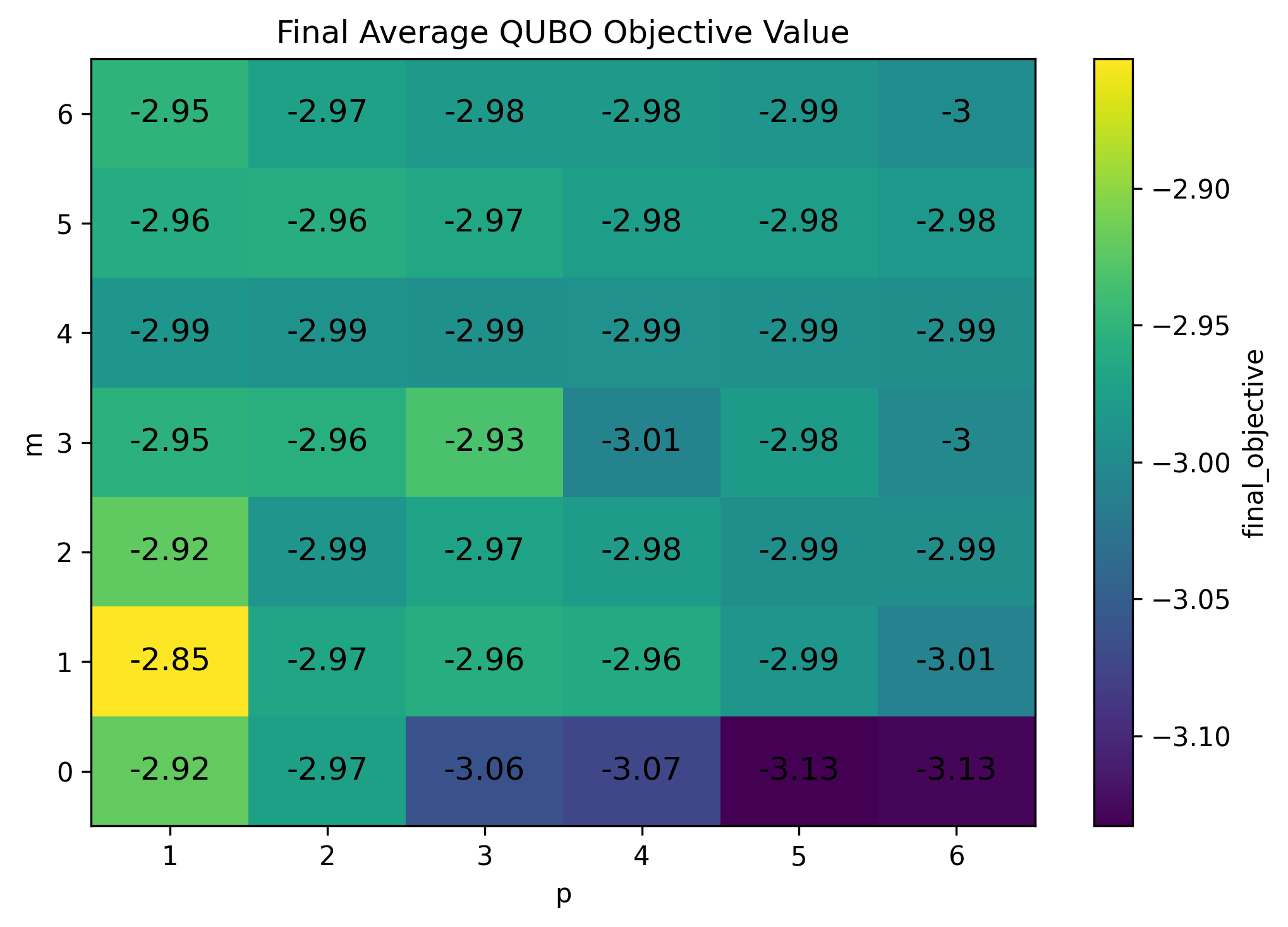}
    \caption{Final average objective value across the $(m,p)$ parameter grid. }
    \label{fig:obj_lcqaoa}
    \end{subfigure}
\hfill
    \begin{subfigure}[t]{0.49\linewidth}
    \centering
    \includegraphics[width=1.0\linewidth]{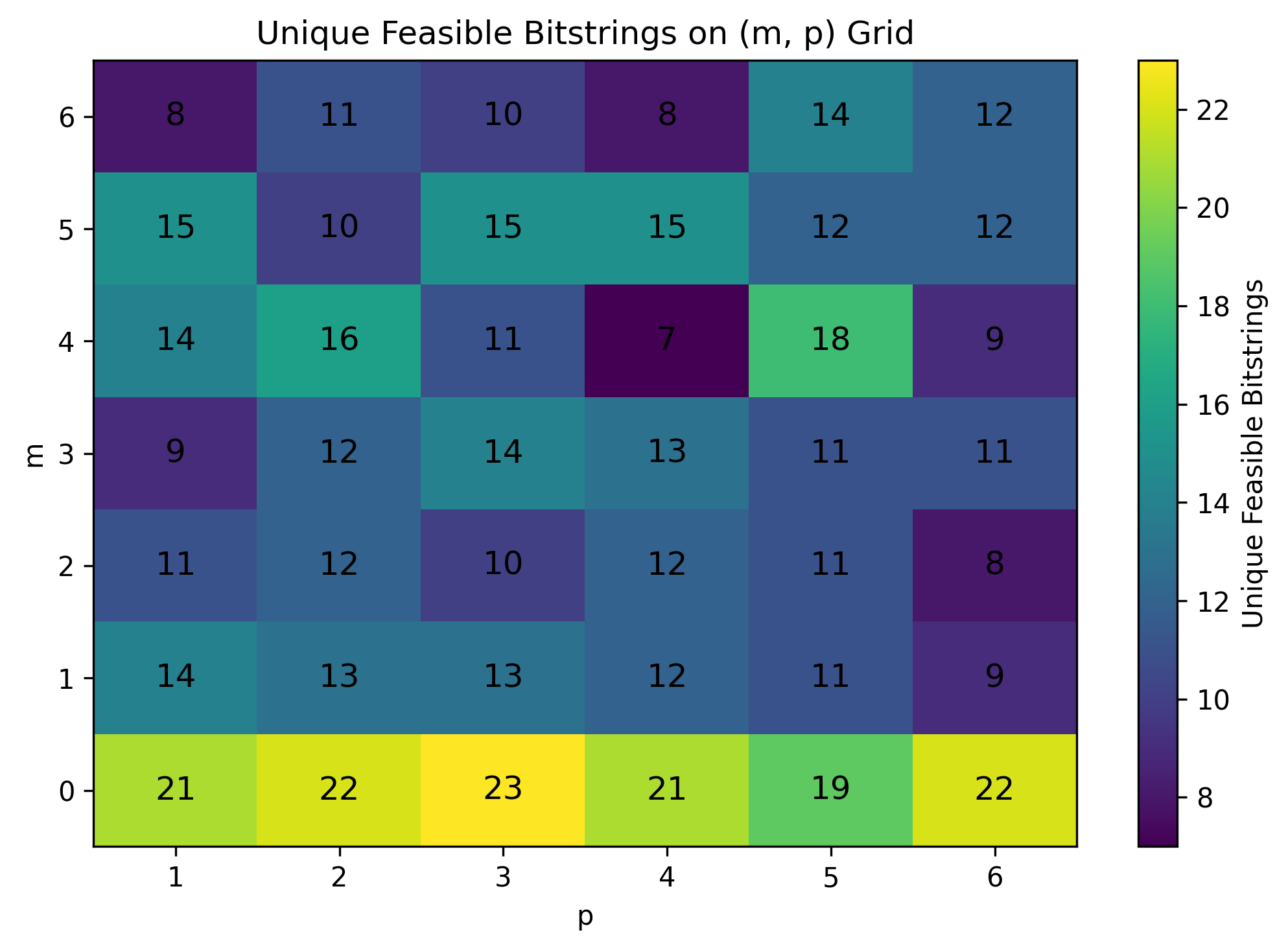}
    \caption{Unique feasible routes samples across the $(m,p)$ parameter grid. }
    \label{fig:heatmap_feas_lcqaoa}
    \end{subfigure}
    \caption{Performance of AQC compressed layers as initial state circuits for LC-QAOA across the $(m,p)$ grid. Results show that LC-QAOA works best without any adiabatic initialization.}
    \label{fig:heatmaps_lcqaoa}
\end{figure*}

Figures~\ref{fig:heatmap_2q_lcqaoa}--\ref{fig:obj_lcqaoa} summarize the behavior of LC-QAOA combined with AQC-compressed prefixes across the $(m,p)$ parameter grid. The results highlight a distinct interaction between compression and variational depth compared to the full-cost QAOA formulation.

The two-qubit depth shows a clear transition as the compression length increases. For small $m$, the depth grows with $p$, reflecting the contribution of additional LC-QAOA layers. However, for $m \geq 4$, the depth becomes nearly constant across all $p$, indicating that the compressed prefix dominates the circuit structure. In this regime, the additional LC-QAOA layers introduce minimal hardware overhead due to their local connectivity.

The classical optimization effort remains primarily governed by the QAOA depth. The number of iterations increases steadily with $p$ across all values of $m$, while variation across $m$ is relatively small. This again suggests that prefix compression mainly influences the initialization rather than the intrinsic difficulty of the variational optimization.

From a feasibility perspective, the uncompressed case ($m=0$) performs particularly well, consistently producing a large number of feasible solutions across different values of $p$. In this setting, the algorithm relies entirely on the LC-QAOA variational layers without any annealing-based initialization. As the compression length increases, the number of feasible samples becomes more variable and generally decreases, indicating that the compressed prefix does not consistently improve feasibility for the LC-QAOA formulation.

The objective values further reveal that performance is relatively flat across the $(m,p)$ grid, especially for larger $m$. While deeper circuits improve the objective in the uncompressed case, this benefit diminishes as compression increases. This behavior can be attributed to the mismatch between the annealing prefix, generated using the full problem Hamiltonian, and the truncated local Hamiltonian used in LC-QAOA.

Overall, these results indicate a tradeoff between hardware efficiency and solution refinement. LC-QAOA with compressed prefixes yields shallow circuits and stable feasibility, but its ability to further improve routing solutions is limited. In contrast, the uncompressed case benefits more from increased variational depth, suggesting that LC-QAOA is most effective when operating directly on the variational landscape rather than relying on compressed annealing prefixes.

\section{Discussion}

This study evaluates hybrid quantum optimization architectures for transportation problems by combining compressed annealing prefixes with variational circuits. Across FLP, VRP, and TSP instances, the results show that approximate quantum compilation (AQC) can reduce circuit depth while preserving access to feasible and high-quality solutions. At the same time, the benefit of compression depends strongly on the interaction between the compressed prefix and the downstream variational ansatz. In this sense, the value of quantum circuit compression for transportation optimization is strongly problem and ans\"atz dependent.

Across these problem classes, the impact of prefix compression on compiled circuit depth is not uniform. In the facility location and vehicle routing cases, the final transpiled two-qubit depth remains irregular across the compression sweep, even in regimes where feasible-solution discovery improves. In contrast, the traveling salesman problem exhibits a clearer reduction in circuit depth as the compression level increases, while maintaining competitive feasible-solution counts.
This distinction highlights an important practical consideration of hardware efficiency, which depends not only by problem size, but also by the structure of the underlying binary encoding. Assignment-based formulations such as the TSP, which impose regular row–column constraints, produce more structured interaction patterns that are more amenable to compression and efficient compilation. In contrast, routing and facility location formulations induce denser and less regular interaction graphs after QUBO mapping, leading to less predictable compilation behavior and limiting the extent to which prefix compression can reduce circuit depth.
From a transportation perspective, this suggests that formulation choice is a central part of hardware-aware quantum optimization. Models with more regular constraint structure may benefit more directly from compilation-based depth reduction techniques, while routing and siting problems may require additional algorithmic or structural modifications to achieve comparable hardware efficiency.

An important practical consideration is that the real-world objective is not simply minimizing an abstract energy function, but generating feasible and diverse candidate solutions under operational constraints. In this context, AQC compression proves useful when paired with standard QAOA. Moderate prefix compression consistently maintains or improves feasible-solution discovery while reducing hardware cost. This suggests that early segments of the annealing trajectory encode useful structural information about routing and assignment feasibility, which can be preserved in a compressed form and effectively refined by subsequent QAOA layers. Therefore, hybrid quantum workflows should be evaluated as formulation-sensitive tools rather than as problem-agnostic solvers. In practical planning settings, the relevant question is not whether a quantum method uniformly reduces hardware cost, but whether it can improve the generation of feasible and decision-relevant candidate solutions under realistic circuit constraints. The present results suggest that moderate prefix compression can be useful in exactly this role. Even when circuit depth is not consistently reduced, compression can still reshape the prepared quantum state in ways that improve feasible-solution discovery

However, the results also show that this benefit does not generalize across all variational architectures. In particular, the LC-QAOA formulation exhibits a markedly different behavior. While LC-QAOA achieves substantial reductions in circuit depth due to its restricted connectivity, its performance does not consistently improve with AQC-based initialization. In many cases, the uncompressed configuration ($m = 0$), corresponding to purely variational initialization, produces comparable or better feasible-solution counts. This indicates that the compressed annealing prefix does not always provide a useful starting point for the LC-QAOA optimization process.

The underlying reason lies in the mismatch between the Hamiltonian used to generate the annealing prefix and the truncated Hamiltonian used in LC-QAOA. In the standard QAOA setting, both the prefix and the variational layers are derived from the same problem Hamiltonian, allowing the prefix to act as a continuation of the same optimization trajectory. In contrast, LC-QAOA modifies the cost Hamiltonian by restricting interactions to local terms, breaking this consistency. As a result, the compressed prefix does not align with the variational landscape explored by LC-QAOA, limiting its ability to improve performance.

This observation highlights an important design principle for hybrid quantum optimization in transportation applications, that improvements in circuit compression, such as reduced two-qubit depth, do not automatically translate into improvements in overall algorithm performance. While AQC can effectively approximate annealing dynamics, its applicability depends on whether the variational algorithm that follows is compatible with the structure of the prepared state. In practical terms, this means that prefix compression should be considered jointly with the choice of variational ans\"atz, rather than as a standalone enhancement.

More broadly, these results reinforce the view of quantum optimization as a candidate-generation tool within transportation analytics pipelines. Shallow circuits enabled by AQC compression are particularly valuable because they produce large numbers of feasible solutions under realistic hardware constraints. For routing and logistics problems, where planners often require multiple viable alternatives rather than a single optimal solution, this tradeoff between depth and feasibility is especially relevant.

At the same time, the findings caution against assuming that improvements in one component of a hybrid quantum algorithm will generalize across architectures. Future work should therefore focus on co-designing initialization strategies and variational circuits that share consistent Hamiltonian structure, enabling more effective integration of compressed annealing dynamics with hardware-efficient optimization layers.

\section{Conclusion}

This paper examined a hardware-aware quantum optimization framework for transportation problems based on compressed adiabatic prefixes and variational circuit tails. Using representative instances of facility location, vehicle routing, and traveling salesman formulations, the results showed that approximate quantum compilation with tensor networks can reduce circuit depth while preserving access to feasible and high-quality candidate solutions. More importantly, the results suggest that quantum methods for constrained transportation problems should be evaluated not only by objective value, but also by their ability to generate feasible, diverse, and decision-relevant alternatives under realistic hardware limits. 

Across the Trotterized annealing experiments, prefix compression produced non-monotonic but often favorable tradeoffs between depth and feasible solution discovery. Moderate compression improved feasible-solution generation in FLP and selected VRP settings, while TSP exhibited clearer depth reductions under stronger compression, indicating that the effectiveness of hardware-aware compression depends strongly on problem formulation and encoding structure. These findings suggest that transportation models with more regular binary interaction patterns may be more amenable to compilation-based depth reduction than routing and siting models with denser or less structured couplings. 

For the VRP hybrid experiments, compressed prefixes followed by standard QAOA provided the most practically useful regime. Moderate compression preserved or improved feasible-solution discovery, and low-depth circuits often retained competitive objective values while producing many feasible routes. This is particularly relevant for transportation planning, where decision-makers frequently benefit more from a set of valid route alternatives than from a single nominally best solution. Because all experiments are conducted on real quantum hardware, the observed tradeoffs directly reflect the practical constraints faced in deploying quantum optimization for transportation systems.

At the same time, the comparison with LC-QAOA showed that the benefits of AQC-based initialization do not automatically transfer across variational architectures. LC-QAOA performed best with no compressed prefix, highlighting that reduced hardware cost alone is not sufficient unless the variational tail remains compatible with the structure encoded by the prefix. This is an important methodological lesson for applied quantum optimization: lighter circuits are valuable only when they preserve the problem information needed to generate feasible and useful plans. 

Overall, these results position hybrid quantum optimization as a candidate-generation component within broader transportation analytics workflows rather than as a standalone solver. In future applications, compressed quantum circuits may be most useful when paired with classical post-processing, simulation, or robustness screening to evaluate candidate routes, assignments, or siting decisions. Future work should therefore focus on extending to larger and more realistic transportation instances, co-design of initialization and variational circuits with consistent Hamiltonian structure, and integrating quantum candidate generation into hybrid decision-support pipelines for logistics and mobility systems.




\bibliographystyle{cas-model2-names}

\bibliography{cas-refs}

\section*{Data Availability}
The code and generated data for the experiments is available at https://github.com/TalhaAzfar/hardware-aware-transport-qaoa/

\section*{Acknowledgments}
This work is funded through the IBM-RPI Future of Computing Research Collaboration. The funder played no role in study design, data collection, analysis and interpretation of data, or the writing of this manuscript. All authors reviewed the final manuscript.

\section*{Declaration of competing interest}
The authors have no competing interests to declare that are relevant to the content of this
article.





\end{document}